\documentclass[prb,letterpaper,aps,superscriptaddress,floatfix,twocolumn,nofootinbib]{revtex4-2}
\pdfoutput=1
\usepackage{graphicx}
\usepackage{amsmath}
\usepackage{amsfonts}
\usepackage{amssymb}
\usepackage[normalem]{ulem}
\usepackage[small,bf]{subfigure}
\usepackage[utf8]{inputenc}
\usepackage{dsfont}
\usepackage{xcolor}
\usepackage{slashed}
\usepackage{soul}
\usepackage{comment}
\usepackage{tikz-cd}
\usepackage{hyperref}
\usepackage{floatpag}
\usepackage{float}
\usepackage{graphicx}
\usepackage{caption}
\usepackage{subcaption}

\newcommand{\beq}{\begin{equation}}
\newcommand{\eeq}{\end{equation}}

\newcommand{\beqa}{\begin{eqnarray}}
\newcommand{\eeqa}{\end{eqnarray}}

\newcommand{\ea}{\end{array}}

\def\eea{\end{eqnarray}}

\def\Tr{ {\rm Tr} }
\def\<{\langle}
\def\>{\rangle}

\usepackage{amsmath}
\usepackage{comment}
\usepackage{amssymb}
\usepackage{amsthm}

\theoremstyle{definition}

\usepackage{color}
\usepackage{tikz-cd}
\usepackage{comment}

\usepackage{tikz}
\usepackage[export]{adjustbox}

\def\[#1\]{%
  \begin{equation}\begin{gathered}#1\end{gathered}\end{equation}%
}

\usepackage[safe]{tipa}
\allowdisplaybreaks

\begin{document}
\title{Topological defects of $2+1$D systems from line excitations in $3+1$D bulk}
\author{Wenjie Ji}
\affiliation{Department of Physics, \mbox{California Institute of Technology, Pasadena, CA, 91125, USA}}
\author{Xie Chen}
\affiliation{Department of Physics, \mbox{California Institute of Technology, Pasadena, CA, 91125, USA}}
\affiliation{Walter Burke Institute for Theoretical Physics and Institute for Quantum Information and Matter, \mbox{California Institute of Technology, Pasadena, CA, 91125, USA}}

\begin{abstract}
The bulk-boundary correspondence of topological phases suggests strong connections between the topological features in a $d+1$-dimensional bulk and the potentially gapless theory on the $(d-1)+1$-dimensional boundary. In $2+1$D topological phases, a direct correspondence can exist between anyonic excitations in the bulk and the topological point defects/primary fields in the boundary $1+1$D conformal field theory. In this paper, we study how line excitations in $3+1$D topological phases become line defects in the boundary $2+1$D theory using the Topological Holography / Symmetry Topological Field Theory framework. We emphasize the importance of ``descendent'' line excitations and demonstrate in particular the effect of the Majorana chain defect: it leads to a distinct loop condensed gapped boundary state of the $3+1$D fermionic $\mathbb{Z}_2$ topological order, and leaves signatures in the $2+1$D Majorana-cone critical theory that describes the transition between the two types of loop condensed boundaries. 
Effects of non-invertible line excitations, such as Cheshire strings, are also discussed in bosonic $3+1$D topological phases and the corresponding $2+1$D critical points.
\end{abstract}

\maketitle
\tableofcontents

\section{Introduction}

\begin{figure}[ht]
    \centering
    \includegraphics[scale=0.55]{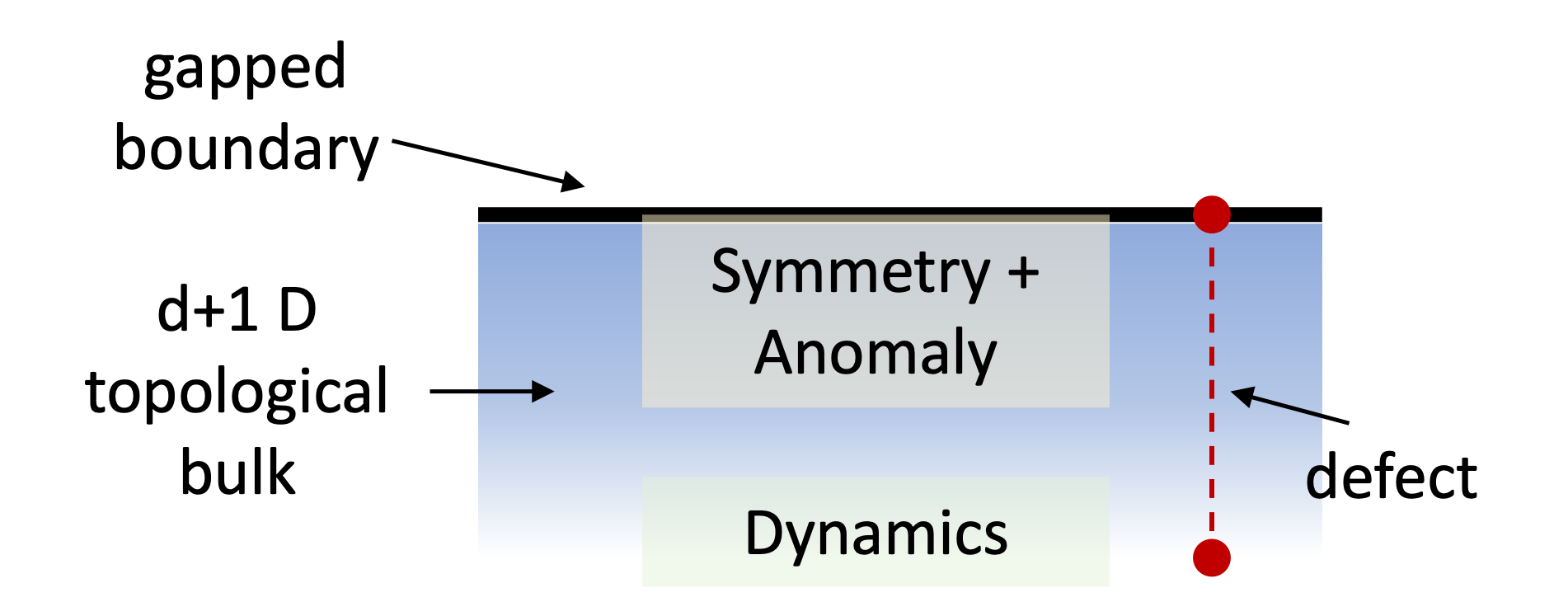}
    \caption{The Topological Holography / Symmetry Topological Field Theory formalism: the $d+1$D topological bulk determines the anomaly of the system; the gapped top boundary determines the symmetry of the system; the bulk and top boundary are held fixed while dynamics takes place at the bottom of the system. Topological excitations in the bulk can tunnel from the top boundary to the bottom boundary and maps the system to a ``topological defect sector''.}
    \label{fig:symTFT}
\end{figure}

The bulk-boundary correspondence is a powerful tool in the study of quantum many-body phases. This power is on full display in the Topological Holography / Symmetry Topological Field Theory (symTFT) framework\cite{Ji2020,kong2020algebraic,gaiotto2021orbifold,Lichtman2021,Freed2023topological,apruzzi2023symmetry,Chatterjee2023symmetry} where a $(d-1)+1$ dimensional quantum theory is realized in a sandwich structure with a $d+1$ dimensional topological theory in the bulk, as shown in Fig.~\ref{fig:symTFT}. The top boundary of the bulk theory is gapped and fixed, while the bottom boundary is left open. With a finite distance between the top and bottom boundary, the sandwich structure effectively represents a $(d-1)+1$ dimensional quantum system. The topological bulk encodes the information of possible symmetries as well as anomalies of the $(d-1)$+1 dimensional quantum theory. A choice of a top gapped boundary explicitly determines the symmetry. The bottom boundary hosts all the local dynamics. A topological excitation inside the bulk can tunnel from the top boundary to the bottom boundary, as indicated by the red dashed line. If the topological excitation (red dot) is condensed at the top boundary, the tunneling process creates excitations only at the bottom boundary, and the excitation is a charge excitation of the $(d-1)+1$ dimensional sandwich. However, if the topological excitation is not condensed, the tunneling process creates excitations at both the top and bottom boundary. It generates a ``topological defect" and maps the sandwich theory to a different topological sector. Because of the similarity between symmetry charges and topological defects, in the following discussion, we often use ``topological defect'' to refer to symmetry charges as well.

This correspondence between topological excitations in the $2+1D$ bulk and topological defects on the boundary is well understood \cite{witten1989quantum,Fuch2002A,Fuch2004B,Fuch2004C,Fuch2005D,Kapustin2010surface,Ji2019,ji2021unified},
 and has been recast in the setting of a $1+1$D sandwich \cite{Lichtman2021,Huang2023topological}. Anyons in the $2+1$D bulk topological theory become point defects in the boundary $1+1$D theory, which may correspond to primary fields if the boundary is in a critical state described by a conformal field theory (CFT). We illustrate this correspondence in section~\ref{sec:1+1} with the example of the doubled-Ising bulk topological order and the critical Ising CFT on the boundary.

In higher dimensions, a similar bulk-boundary correspondence exists\cite{Chen2016,Ji2020,ji2023boundary}, but is much less well understood. For $3+1$D gapped topological phases (discrete gauge theories), we usually think of the topological excitations as consisting of gauge charge point excitations and gauge flux loop excitations. However, recent developments suggest that one needs to consider a 2-category structure of the topological excitations\cite{Etingof2009fusion, Kapustin2010topological,Kong2014braided,Kong2015boundarybulk,Douglas2018fusion,Kong2020,Johnson-Freyd2022,Lakshya2023,Bartsch2023higher}. Physically speaking, this means that the fundamental objects are line excitations. Point excitations are domain walls on top of the (trivial or nontrivial) line excitations. Moreover, besides the elementary flux loop excitations, we need to consider ``descendent'' line excitations which are gapped 1D states formed by the gauge charges. 

\begin{figure}[ht]
    \centering
    \includegraphics[scale=0.65]{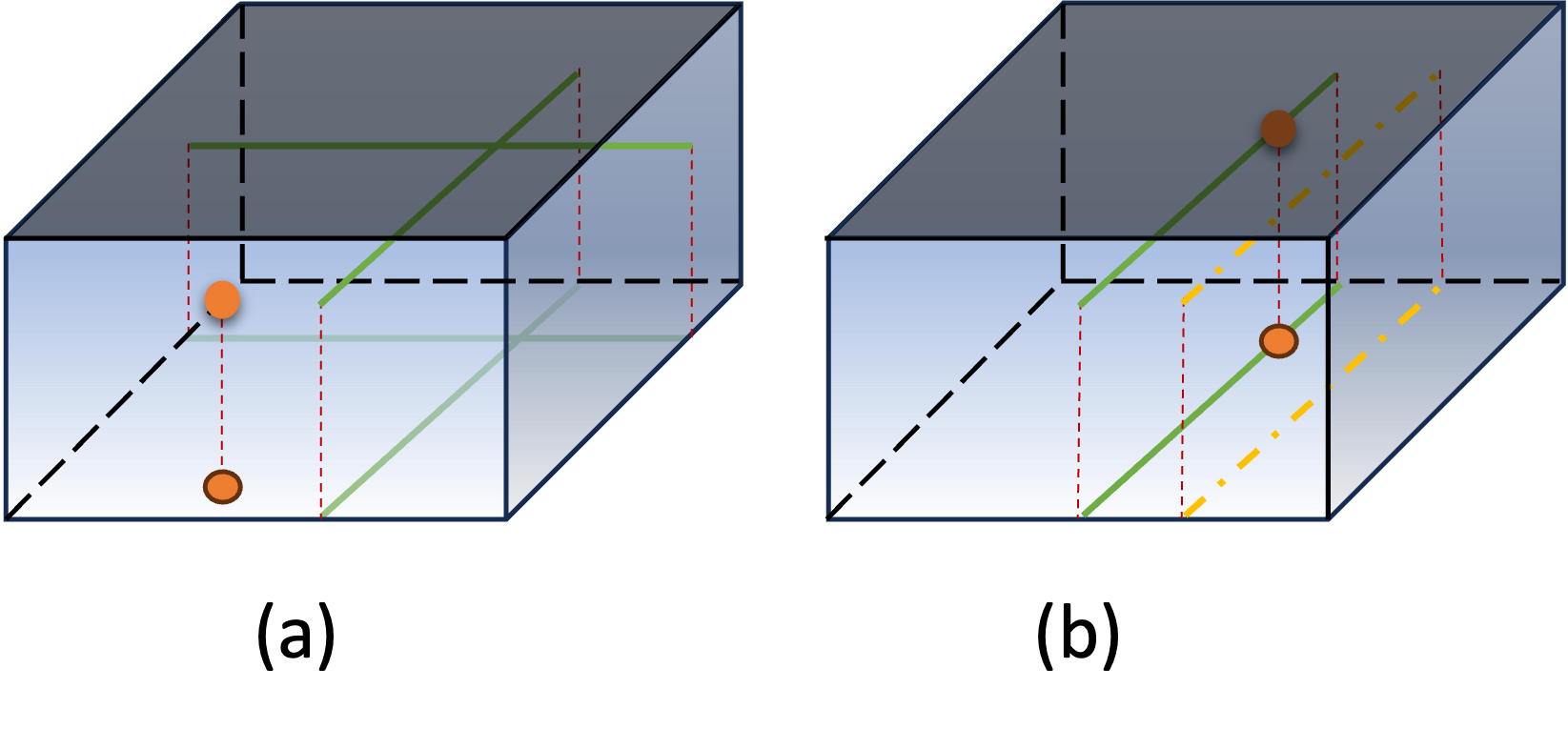}
    \caption{A $2+1$D ``sandwich'' structure with $3+1$D discrete gauge theory in the bulk and gapped boundary condition at the top. Different topological sectors can be obtained by tunneling (a) gauge flux loop excitations (green lines) and gauge charge point excitations (orange dots) or (b) elementary flux loop excitations (green lines) and descendent line excitations (dashed yellow lines) with possible gauge charge domain walls (orange dots) on top.}
    \label{fig:2+1defect}
\end{figure}

Such a change in perspective on the topological excitations in the $3+1$D bulk leads to a corresponding change in perspective on the defects in the $2+1$D boundary theory. As illustrated in Fig.~\ref{fig:2+1defect}, when we think of topological excitations as gauge charges and gauge fluxes, the corresponding defects are point charges and flux lines (for example around nontrivial cycles in the $x$ and $y$ directions). If we think of topological excitations using the 2-category structure, the corresponding defects are line defects with possible domain walls on top. The line defect may come from flux loop, or descendent line excitations, or their composites. 

In this paper, we explore the connection between the two perspectives and study the implications of the 2-category structure of topological excitations in the $3+1$D bulk on the line defects in the $2+1$D boundary. In particular, we pay special attention to the role played by descendent line excitations and demonstrate in section~\ref{sec:fZ2},\ref{sec:Z2}, \ref{sec:Z2Z2} through explicit examples what kinds of topological sectors are induced by the descendent defects and their potential effect on the boundary theory. 

In section~\ref{sec:fZ2}, we consider the $2+1$D superconducting systems as the boundary of the $3+1$D $\mathbb{Z}_2$ gauge theory with fermionic gauge charge. The descendent line excitation in the bulk theory is the 1D Majorana chain formed by the fermionic gauge charge. We see that due to the existence of the Majorana chain excitation, there are two distinct types of loop-condensed gapped boundary states of the $3+1$D bulk corresponding to the $2+1$D trivial and $p+ip$ superconducting states respectively. The transition between the two states involves a $2+1$D Majorana cone. Through the sandwich construction, we see how different topological sectors of the Majorana cone combine according to rules given by the condensate on the top gapped boundary such that the whole sandwich describes a modular invariant theory in $2+1$D. We future calculate the zero point energy of different topological sectors and identify how the scaling of their difference embodies the competition between the two loop condensates in a quantitative way.


In section~\ref{sec:Z2}, we consider the $2+1$D spin system with global 0-form $\mathbb{Z}_2$ symmetry as the boundary of the $3+1$D $\mathbb{Z}_2$ gauge theory with bosonic charge. The descendent line excitation in the bulk theory is the Cheshire string -- a 1D condensate of the gauge charge. We discuss how the non-invertible fusion rule of the excitation is manifested in the projection operation induced in the bulk by the excitation and its implication on the topological sectors of the boundary theory.

In section~\ref{sec:Z2Z2}, we consider the $2+1$D spin system with anomalous 0-form $\mathbb{Z}_2\times \mathbb{Z}_2$ symmetry as the boundary of the $3+1$D twisted $\mathbb{Z}_2\times \mathbb{Z}_2$ gauge theory with bosonic charge. This theory is special in that the flux loop excitations in the bulk have to be charge condensates, therefore non-invertible. We discuss the topological sectors induced by the various line defects and the implications on the possible transition point between the trivial and nontrivial $2+1$D bosonic SPT phases with $\mathbb{Z}_2$ symmetry.


\section{Review: $1+1$D critical Ising chain}
\label{sec:1+1}

In this section, we review how the topological defects in the $1+1$D critical Ising chain (spin system with global $\mathbb{Z}_2$ symmetry) are related to topological excitations in the $2+1$D doubled-Ising topological state and how the modular invariant partition function of the critical theory arises naturally from the sandwich picture. 

\begin{figure}[ht]
    \centering
    \includegraphics[scale=0.7]{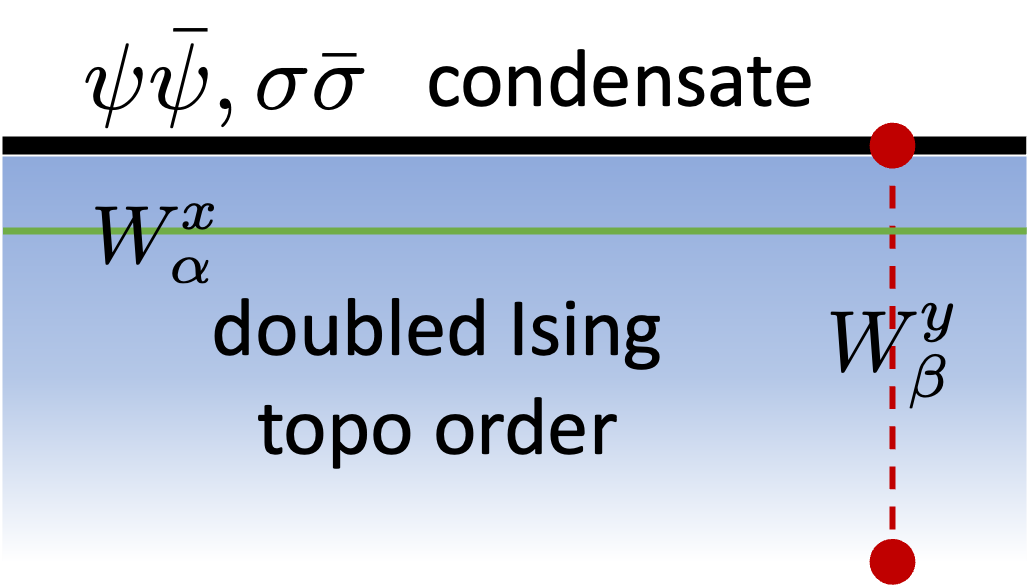}
    \caption{A $1+1$D ``sandwich'' structure with the $2+1$D doubled-Ising topological order in the bulk. The top boundary is gapped by condensing $\psi\bar{\psi}$ and $\sigma\bar{\sigma}$. Different topological sectors are labeled by the eiganvalues of the $x$-direction string operators of anyon $\alpha$, $W^x_{\alpha}$, and can be obtained by tunneling anyon $\beta$ in the $y$ direction with string operator $W^y_{\beta}$ starting from the trivial sector.}
    \label{fig:Ising}
\end{figure}

Consider a symTFT sandwich as shown in Fig.~\ref{fig:Ising} with the $2+1$D doubled-Ising topological order in the bulk. There are nine types of anyons in the doubled-Ising topological order coming from a chiral Ising topological state and its time reversal counterpart,
\begin{equation}
\alpha \in \{\mathbb{I}, \sigma, \psi\} \times \{\mathbb{I}, \bar{\sigma}, \bar{\psi}\}.
\end{equation}
The top boundary of the sandwich is gapped out by the condensation of $\psi\bar{\psi}$ and $\sigma\bar{\sigma}$. Following the convention in ..., we  denote the gapped boundary state as a nine-dimensional vector with integer entries,
\begin{equation}
(1 ~0~  0~  0~ 1~ 0~ 0~ 0~  1),
\label{eq:Ising_t}
\end{equation}
where the nine-dimensions are labeled by anyons $\{\mathbb{I},\psi,\sigma,\bar{\psi},\psi\bar{\psi},\sigma\bar{\psi},\bar{\sigma},\psi\bar{\sigma},\sigma\bar{\sigma}\}$.
If the condensate has no defect, the horizontal string operators satisfy $W^x_{\psi\bar{\psi}} = W^x_{\sigma\bar{\sigma}}=1$. The horizontal string operators of $\sigma$ and $\psi$ become the symmetry operators of the sandwich which satisfy the fusion rule of
\begin{equation}
W^x_{\sigma}W^x_{\sigma} = W^x_{\psi}+W^x_{\mathbb{I}},\  W^x_{\psi}W^x_{\psi} = W^x_{\mathbb{I}}.
\end{equation}
$W^x_{\psi}$ represents the $\mathbb{Z}_2$ symmetry of the Ising chain while $W^x_{\sigma}$ represents the non-invertible duality symmetry.

The bottom boundary can be tuned to the Ising critical point as long as neither $\sigma\bar{\sigma}$ or $\psi\bar{\psi}$ condenses. The critical theory has nine topological sectors corresponding to the nine anyons in the bulk. If we start from the trivial sector labeled by $W^x_{\psi} = W^x_{\bar{\psi}} = 1$, $W^x_{\sigma} = W^x_{\bar{\sigma}} = \sqrt{2}$, we can reach the other sectors by tunneling different anyons from the top boundary to the bottom boundary with $W^y_{\beta}$. The Wilson line $W^y_{\beta}$ lands on the boundary as a topological line along the time direction. The eigenvalues of $W_{\alpha}^x$ with $\alpha =\psi,\bar\psi,\sigma,\bar\sigma$ in the sector labeled by anyon $\alpha$ are determined by the Verlinda algebra \cite{verlinde1988fusion}
\begin{align}
    W_\alpha^x W_\beta^y |0\rangle =\frac{S_{\beta\alpha}}{S_{1\alpha}}W_\beta^y |0\rangle,
\end{align}
where $|0\rangle$ represents the trivial sector of the bottom boundary, and $S_{ij}$ are matrix elements of the modular $S$ matrix of the bulk topological order. Each sector then corresponds to a conformal tower in the Ising critical CFT. Table I lists all nine sectors labeled by the anyon being tunneled, the eigenvalue of $W^x_{\psi}$, $W^x_{\bar{\psi}}$,$W^x_{\sigma}$, $W^x_{\bar{\sigma}}$, and the conformal weight of the corresponding primary field in the critical Ising CFT. The conformal weight of each defect can be calculated from the energy and momentum of ground state when the defect is inserted.

\begin{table}[th]
    \centering
    \begin{tabular}{||c|c|c|c|c|c||}
        Anyon & $W^x_{\psi}$ & $W^x_{\bar{\psi}}$ & $W^x_{\sigma}$ & $W^x_{\bar{\sigma}}$ & $\Delta$\\
         $\mathbb{I}$ & $1$ & $1$ & $\sqrt{2}$ & $\sqrt{2}$ & $(0,0)$\\
         $\sigma\bar{\sigma}$ & $-1$ & $-1$ & $0$ & $0$ & $(1/16,1/16)$\\
         $\psi\bar{\psi}$ & $1$ & $1$ & $-\sqrt{2}$ & $-\sqrt{2}$ & $(1/2,1/2)$ \\
         $\sigma$ & $-1$ & $1$ & $0$ & $\sqrt{2}$ & $(1/16,0)$\\
         $\bar{\sigma}$ & $1$ & $-1$ & $\sqrt{2}$ & $0$ & $(0,1/16)$\\
         $\psi$ & $1$ & $1$ & $-\sqrt{2}$ & $\sqrt{2}$ & $(1/2,0)$\\
         $\bar{\psi}$ & $1$ & $1$ & $\sqrt{2}$ & $-\sqrt{2}$ & $(0,1/2)$\\
         $\sigma\bar{\psi}$ & $-1$ & $1$ & $0$ & $-\sqrt{2}$ & $(1/16,1/2)$\\
         $\psi\bar{\sigma}$ & $1$ & $-1$ & $-\sqrt{2}$ & $0$ & $(1/2,1/16)$\\
    \end{tabular}
    \caption{Topological sectors in the $1+1$ Ising critical CFT: anyon label; eigenvalues of the topological symmetry operators ($W^x_{\psi}$, $W^x_{\bar{\psi}}$, $W^x_{\sigma}$, $W^x_{\bar{\sigma}}$); the conformal dimensions of the primary field.}
    \label{table:1DIsing}
\end{table}

Not all nine sectors show up in the partition function of the critical Ising chain. Only certain combination is allowed to satisfy the requirement of modular invariance. It is well known that the modular invariant partition function of the critical Ising chain is given by
\begin{equation}
Z = Z_{\mathbb{I}} + Z_{\psi\bar{\psi}} + Z_{\sigma\bar{\sigma}}.
\label{eq:Ising_Z}
\end{equation}
This can be easily derived in the sandwich picture. On two-dimensional torus, the doubled-Ising topological model has a nine-dimensional ground space. A basis can be chosen for the ground space as the common eigenstates of $W^x_{\psi}$, $W^x_{\sigma}$, $W^x_{\bar{\psi}}$, $W^x_{\bar{\sigma}}$. These states are the ``Minimally Entangled States'' with respect to the entanglement cuts along the $x$ direction and can be obtained from the state with $W^x_{\psi}=W^x_{\bar{\psi}}=1$, $W^x_{\sigma}=W^x_{\bar{\sigma}}=\sqrt{2}$ by tunneling anyon $\beta$ in the $y$ direction using $W^y_{\beta}$. Therefore, the Minimally Entangled basis states are labeled by $\alpha$. On the other hand, the nine sectors of the $1+1$ critical theory are also labeled by $\alpha$ and they transform under the space-time (2D) modular transformations in exactly the same way the nine Minimally Entangled states of the $2+1$D doubled-Ising topological state transform under the spatial (2D) modular transformation. In this basis, the modular transformations are generated by
\begin{equation}
\begin{array}{ll}
S_{\text{Ising}} = s \otimes s^{\dagger}, & T_{\text{Ising}} = t \otimes t^{\dagger}, \\ 
s=\frac{1}{2}\begin{pmatrix} 1 & 1 & \sqrt{2} \\ 1 & 1 & -\sqrt{2} \\ \sqrt{2} & -\sqrt{2} & 0 \end{pmatrix}, & t=\begin{pmatrix} 1 & 0 & 0 \\ 0 & -1 & 0 \\ 0 & 0 & e^{i2\pi/16} \end{pmatrix}.
\end{array}
\end{equation}
The nine sectors are hence ``modular covariant''. To get a modular invariant partition function of the $1+1$D critical theory, we can take an ``inner product'' between the nine-dimensional vector of topological sectors of the critical theory and the nine-dimensional vector corresponding to the top gapped boundary given in Eq.~\ref{eq:Ising_t}. The vector corresponding to the top gapped boundary is modular invariant (has eigenvalue $1$ under both $S_{\text{Ising}}$ and $T_{\text{Ising}}$). Therefore, the inner product, which gives rise to the partition function in Eq.~\ref{eq:Ising_Z}, is modular invariant. In the next section, we are going to see how similar construction gives rise to a modular invariant partition function of the $2+1$D Majorana cone on the boundary of the $3+1$D fermionic toric code.


\section{$2+1$D superconducting systems}
\label{sec:fZ2}

In the symTFT framework, $2+1$D fermionic system with 0-form $\mathbb{Z}_2$ symmetry (with or without coupling to a dynamical $\mathbb{Z}_2$ gauge field) can be realized on the boundary of the $3+1$D fermionic Toric Code ($3+1$D $\mathbb{Z}_2$ gauge theory with fermionic charge). In this section, we are going to establish a similar correspondence between the modular invariant \footnote{when coupled to $\mathbb{Z}_2$ gauge field, the bosonic system is invariant under the $SL(3,\mathbb{Z})$ modular transformations; when not coupled to the gauge field, the fermionic partition functions with different boundary conditions form a unitary representation under the modular group $SL(3,\mathbb{Z})$ with a $\mathbb{Z}_2^F$ extension.} partition function of the $2+1$D critical superconducting state and the topological sectors in the $3+1$D fermionic Toric Code state. The anyon basis which labels the topological sectors given a 2+1D bulk is now generalized to the \emph{minimally entangled basis}\cite{Zhang2012} given a 3+1D bulk. Most interestingly, we are going to see the special role played by descendent line excitations / defects in this correspondence. 

\subsection{$3+1$D fermionic Toric Code: bulk}


Let us first discuss the bulk properties of the $3+1$D fermionic Toric Code. As a $\mathbb{Z}_2$ gauge theory, the $3+1$D fermionic toric code contains a point excitation of fermionic gauge charge and a loop excitation of gauge flux. Moreover, there is a descendent line excitation of Majorana chain formed by the fermionic gauge charge. On a 3D torus, there is an eight-fold degenerate ground space (of three logical qubits) where the fermionic string operators $W^x_f$, $W^y_f$, $W^z_f$ and the $\mathbb{Z}_2$ flux membrane operators $V^{yz}_m$, $V^{zx}_m$, $V^{xy}_m$ act as logical Pauli operators
\begin{equation}
\begin{array}{lll}
W^x_f \sim Z_1, & W^y_f \sim Z_2, & W^z_f \sim Z_3, \\
V^{yz}_m \sim X_1, & V^{zx}_m \sim X_2, & V^{xy}_m \sim X_3. \\
\end{array}
\label{eq:fTC_WV}
\end{equation}
WLOG, we assume that the state with no flux ($W^x_f = W^y_f = W^z_f = 1$) corresponds to anti-periodic boundary condition in all three directions for the fermionic charges. 

Let us focus on the Majorana chain excitation $M$ and see what kind of logical operation is induced by sweeping a Majorana chain across, for example, the $yz$ plane with the membrane operator $V^{yz}_M$. Since the Majorana chain realizes an invertible phase with $\mathbb{Z}_2$ classification, the Majorana line excitation has a $\mathbb{Z}_2$ fusion rule,
\begin{equation}
M \times M = \mathcal{I}.
\end{equation}
The corresponding membrane operator $V^{yz}_M$ implements a unitary transformation on the ground space of the fermionic toric code that squares to identity. To see what kind of unitary this is, we notice that a Majorana chain has odd fermion parity with periodic boundary and even fermion parity with anti-periodic boundary condition. Therefore, when a Majorana chain in the $y$ direction is pumped by the membrane operator $V^{yz}_{M}$, it tunnels a charge in the $z$ direction depending on whether there is a $\pi$ flux in the $y$ direction. That is, on the three logical qubits specified above, it implements a controlled-$Z$ gate between the second and third logical qubits. This result was derived in 
Ref.~\cite{Barkeshli2024}. Similar relation holds for Majorana chains in other directions. That is
\begin{equation}
V^{yz}_M \sim \text{CZ}_{23}, \ V^{zx}_M \sim \text{CZ}_{31}, \ V^{xy}_M \sim \text{CZ}_{12},
\end{equation}

The eight-fold degenerate ground space transforms unitarily under the modular transformations of the 3D torus. The explicit form of the modular transformations can be derived following the general formalism developed in Ref.~\cite{Wang2017}, which we briefly review for the fermionic Toric Code case in appendix~\ref{sec:WW}. In the basis specified above, the generators of the modular transformations take the form
\begin{equation}
S = \begin{pmatrix} 1 & 0 & 0 & 0 & 0 & 0 & 0 & 0 \\ 0 & 0 & 0 & 0 & 1 & 0 & 0 & 0\\ 0 & 1 & 0 & 0 & 0 & 0 & 0 & 0 \\ 0 & 0 & 0 & 0 & 0 & 1 & 0 & 0 \\ 0 & 0 & 1 & 0 & 0 & 0 & 0 & 0 \\ 0 & 0 & 0 & 0 & 0 & 0 & 1 & 0 \\ 0 & 0 & 0 & 1 & 0 & 0 & 0 & 0 \\ 0 & 0 & 0 & 0 & 0 & 0 & 0 & 1 \end{pmatrix}, 
T = \begin{pmatrix} 0 & 1 & 0 & 0 & 0 & 0 & 0 & 0 \\ 1 & 0 & 0 & 0 & 0 & 0 & 0 & 0\\ 0 & 0 & 1 & 0 & 0 & 0 & 0 & 0 \\ 0 & 0 & 0 & 1 & 0 & 0 & 0 & 0 \\ 0 & 0 & 0 & 0 & 0 & 1 & 0 & 0 \\ 0 & 0 & 0 & 0 & 1 & 0 & 0 & 0 \\ 0 & 0 & 0 & 0 & 0 & 0 & 1 & 0 \\ 0 & 0 & 0 & 0 & 0 & 0 & 0 & 1 \end{pmatrix}. 
\label{eq:fTC_MT}
\end{equation}
There are two modular invariant (eigenvalue-1) vectors under both $S$ and $T$
\begin{align}
v_m = &\begin{pmatrix} 1 & 1 & 1 & 1 & 1 & 1 & 1 & 1\end{pmatrix}, \\
v_{mM} =& \begin{pmatrix} 1 & 1 & 1 & 1 & 1 & 1 & 1& -1 \end{pmatrix},
\end{align}
and a vector that is invariant (eigenvalue-1) under $S$ and $T^2$
\begin{align}
v_f = \begin{pmatrix} 1 & 0 & 0 & 0 & 0 & 0 & 0 & 0\end{pmatrix}.
\end{align}
The subscripts come from the fact that $v_m$ is the common eigenvalue-1 eigenstate of $V^{xy}_m$, $V^{yz}_m$, $V^{zx}_m$, $v_{mM}$ is the common eigenvalue-1 eigenstate of $V^{xy}_mV^{xy}_M$, $V^{yz}_mV^{yz}_M$, $V^{zx}_mV^{zx}_M$, and $v_{f}$ is the common eigenvalue-1 eigenstate of $W^{x}_f$, $W^{y}_f$, $W^z_f$. We are going see that these three vectors correspond to three different types of gapped boundaries of the 
$3+1$D fermionic toric code where the $m$ flux loop, the $m$ flux loop bound together with an $M$ Majorana chain and the $f$ fermionic charge condenses respectively. 

\subsection{$3+1$D fermionic Toric Code: gapped boundaries}

The three vectors listed above correspond to the following gapped boundadries of the $3+1$D fermionic toric code: $v_m$ -- a condensation of flux loop, $v_{mM}$ -- a condensation of flux loop and Majorana chain composite and $v_f$ -- a condensation of fermion charge (paired with a physical fermion)\cite{wen2024topological,huang2024fermionic,Ji2024invertible}.

\begin{figure}[ht]
    \centering
    \includegraphics[scale=0.45]{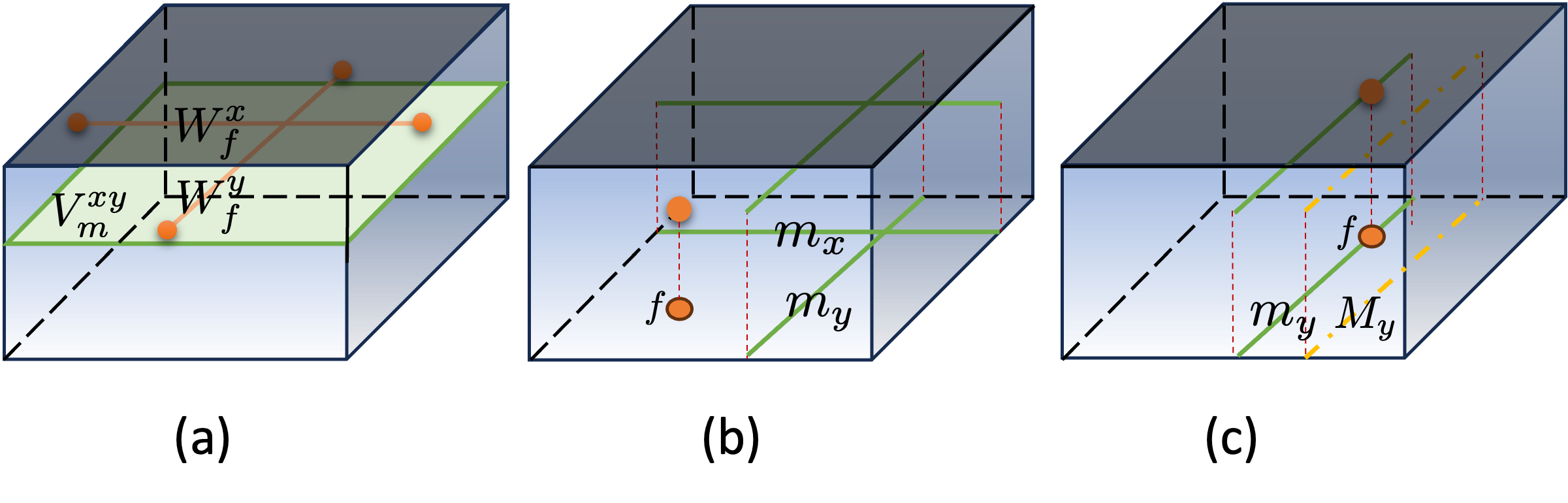}
    \caption{In the $2+1$D sandwich structure with $3+1$D fermionic Toric Code in the bulk, (a) different topological sectors are labeled by the eigenvalues of string and membrane logical operators ($W^x_f$, $W^y_f$, $V^{xy}_m$) parallel to the boundary (i.e., the MES basis). Different sectors can be obtained by (b) tunneling gauge charges ($f$) and gauge flux loops ($m_x$, $m_y$) in the $z$ direction, or (c) tunneling gauge flux loops ($m_y$) and Majorana chains ($M_y$) with possible gauge charge ($f$) domain walls on top.}
    \label{fig:fZ2}
\end{figure}

In order to identify the gapped boundaries of the 3+1D topological order through vectors of partition functions, the anyon basis for the vector is generalized to the \emph{minimally entangled state (MES) basis}\cite{Zhang2012}.
Thus, to see this correspondence, let us switch to the MES basis of the ground space, which are common eigenvectors of the logical operators along the entanglement cut. Let us choose the entanglement cut to be across an $xy$ plane. The Minimally Entangled States are then eigenstates of $W^x_f$, $W^y_f$ and $V^{xy}_m$, as shown in Fig.~\ref{fig:fZ2} (a) (Fig.~\ref{fig:fZ2} shows a system with open boundary while here we consider a system with periodic boundary conditions). Since $W^z_f$ and $V^{xy}_m$ correspond to $Z$ and $X$ operations respectively on the third logical qubit, this basis change corresponds to a Hadamard transformation on the third logical qubit and the three vectors are mapped to
\begin{align}
\tilde{v}_m = \begin{pmatrix} 1 & 0 & 1 & 0 & 1 & 0 & 1 & 0\end{pmatrix}, \\
\tilde{v}_{mM} = \begin{pmatrix} 1 & 0 & 1 & 0 & 1 & 0 & 0 & 1 \end{pmatrix}, \\
\tilde{v}_f = \begin{pmatrix} 1 & 1 & 0 & 0 & 0 & 0 & 0 & 0\end{pmatrix}.
\label{eq:fTC_gapped_boundaries}
\end{align} 

Now we physically cut the $3+1$D system with top and bottom boundaires in the $z$ direction while keeping the periodic boundary conditions in the $x$ and $y$ directions. The eight ground states on the 3D torus now become the eight topological sectors labeled by eigenvalues of $W^x_f$, $W^y_f$ and $V^{xy}_m$. Each of these operators can have $\pm 1$ eigenvalues. Starting from the sector with $W^x_f = W^y_f = V^{xy}_m = 1$, the other sectors can be reached by applying $V^{yz}_m$, $V^{zx}_m$ to tunnel $\pi$ fluxes in the $z$ direction and applying $W^z_f$ to tunnel a fermionic charge in the $z$ direction, as shown in Fig.~\ref{fig:fZ2} (b). The topological sectors are hence labeled by anti-periodic / periodic boundary condition in the x direction, anti-periodic / periodic boundary condition in the y direction, and even / odd total fermion parity on each boundary. 

With this interpretation, we see that $\tilde{v}_m$ corresponds to the superposition of the no-flux-even-charge sector, the $x$-flux-even-charge sector, the $y$-flux-even-charge sector and the $x$-flux-$y$-flux-even-charge sector. Therefore, it describes the boundary state of the $m$ flux loop condensate.
The $\tilde{v}_f$ vector corresponds to the superposition of the no-flux-even-charge sector and the no-flux-odd-charge sector. Therefore, it describes the boundary state of the fermionic charge condensate. 

Let us now see why the $\tilde{v}_{mM}$ vector describes a boundary where the composite of $m$ flux loop and $M$ Majorana chain condenses. The $\tilde{v}_{mM}$ vector corresponds to the superposition of the no-flux-even-charge sector, the $x$-flux-even-charge sector, the $y$-flux-even-charge sector and the $x$-flux-$y$-flux-ODD-charge sector. The only difference from the $\tilde{v}_{m}$ vector is that the sector with both $x$ and $y$ fluxes is fermion parity odd. This is the consequence of tunneling a Majorana chain in the same direction when a flux loop is tunneled, as shown in Fig.~\ref{fig:fZ2} (c). The effect of tunneling a Majorana chain along $x$ or $y$ in the $z$ direction is to tunnel a charge in the $z$ direction when the flux is $\pi$ (the boundary condition is periodic) in $x$ or $y$ direction. Therefore, tunneling the composite of $m$ and $M$ along $x$ has the same effect as tunneling only $m$ along $x$; tunneling the composite of $m$ and $M$ along $y$ has the same effect as tunneling only $m$ along $y$; however, tunneling the composite of $m$ and $M$ along $x$ after tunneling the composite along $y$ (or in the opposite order) differs from tunneling only $m$ along $x$ after tunneling only $m$ along $y$ (or in the opposite order) by the tunneling of a charge, since the second Majorana chain sees a $\pi$ flux.

Note that not all gapped boundary types of the $3+1$D fermionic Toric Code are captured using this method. In particular, there should be 16 different types of loop condensed boundary corresponding to the 16-fold-way of $2+1$D superconductors. The computation we do here starting with a $T^3$ torus geometry only captures two of them.

\subsection{$2+1$D critical Majorana cone}

Having established the correspondence between the three vectors $\tilde{v}_m$, $\tilde{v}_{mM}$ and $\tilde{v}_f$ and the gapped boundary conditions of $3+1$D fermionic Toric Code, we are going to use them to build the $2+1$D sandwich structure for $2+1$D superconducting systems.

Consider the sandwich structure as shown in  Fig.~\ref{fig:phase_diagram}. If the top boundary of the sandwich is gapped by condensing the fermionic charge (paired with another physical fermion, corresponding to vector $\tilde{v}_f$), the relevant symmetry of the sandwich is a $0$-form $\mathbb{Z}_2$ symmetry given by the gauge flux membrane operator in the bulk $V^{xy}_m$ stretching parallel to the boundaries. The sandwich realizes a $2+1$D fermionic system with $\mathbb{Z}_2$ fermion parity symmetry. If the top boundary is gapped by condensing either type of the flux loop (with or without the decoration of the Majorana chain, corresponding to vector $\tilde{v}_m$ or $\tilde{v}_{mM}$), the relevant symmetry of the sandwich is a $1$-form $\mathbb{Z}_2$ symmetry generated by the string operator of the fermion $W_f$ in the $xy$ plane. The sandwich realizes a $2+1$D system with $1$-form $\mathbb{Z}_2$ symmetry. When the $1$-form symmetry is spontaneously broken, the system becomes topologically ordered with a fermionic quasi-particle. Therefore, in this case the sandwich represents a fermionic superconducting system coupled to a dynamical $\mathbb{Z}_2$ gauge field. 

If the bottom boundary is also in one of the three gapped states, the whole sandwich is gapped. The gapped phase realized in each case are listed in Table~\ref{table:fZ2_gapped}. The ground state degeneracy of the gapped phase can be obtained by taking the inner product between the top and bottom vectors.

\begin{table}[th]
    \centering
    \begin{tabular}{||c|c|c||}
        Top & Bottom & Gapped Phase \\[2pt]

         $\tilde{v}_f$ & $\tilde{v}_f$ & trivial superconductor*\\
         $\tilde{v}_f$ & $\tilde{v}_m$ & trivial superconductor\\
         $\tilde{v}_f$ & $\tilde{v}_{mM}$ & $p+ip$ superconductor \\
         $\tilde{v}_{m}$ & $\tilde{v}_{m}$ & $2+1$D Toric Code \\
         $\tilde{v}_{m}$ & $\tilde{v}_{mM}$ & chiral Ising topo order \\
         $\tilde{v}_{mM}$ & $\tilde{v}_{mM}$ & $2+1$D Toric Code \\[1pt]
    \end{tabular}
    \caption{Gapped phases in the $2+1$D sandwich structure with $3+1$D fermionic Toric Code in the bulk. The top and bottom boundaries are gapped by condensing fermion charge ($\tilde{v}_f$), condensing flux loop ($\tilde{v}_m$) or condensing Majorana decorated flux loop ($\tilde{v}_{mM}$). *: there is an extra $\mathbb{Z}_2$ symmetry in this configuration which is spontaneously broken in the ground state.}
    \label{table:fZ2_gapped}
\end{table}

Now let us consider the more interesting case of a gapless sandwich. Suppose that the top boundary is in one of the three gapped states given by $\tilde{v}_f$, $\tilde{v}_m$ and $\tilde{v}_{mM}$. The bottom boundadry is tuned to the critical point between $\tilde{v}_m$ and $\tilde{v}_{mM}$. The phase transition realized is then between a trivial superconductor and a non-trivial $p+ip$ superconductor (either not coupled or coupled to a dynamical $\mathbb{Z}_2$ gauge field). It is well known that the critical point can be realized as a Majorana cone (either not coupled or coupled to a dynamical $\mathbb{Z}_2$ gauge field), which can be modeled at low energy by the Hamesiltonian
\begin{equation}
H = \frac{1}{(2\pi)^2} \int d^2\mathbf{r} \lambda(\mathbf{r})^T \left(-i\sigma_3\partial_x - i\sigma_1\partial_y\right) \lambda(\mathbf{r}).
\end{equation}
where $\sigma_{1,2,3}$ are the Pauli matrices and $\lambda(\mathbf{r})$ is a two-component real fermionic field.
Let us see how a modular invariant partition function can be derived using the sandwich structure following a similar procedure as illustrated in section~\ref{sec:1+1} for the $1+1$D critical Ising chain.

The $2+1$D Majorana cone contains eight sectors, labeled by fermion parity even (e) / odd (o), anti-periodic (ap) / periodic (p) boundary condition in the x direction, and anti-periodic (ap) / periodic (p) boundary condition in the y direction. Correspondingly, there are eight parts of its partition function, corresponding to the eight Minimally Entangled States,
\begin{equation}
\arraycolsep=1.4pt\def\arraystretch{1.5}
\begin{array}{cccc}
Z^{MC}_{e,ap,ap}, & Z^{MC}_{o,ap,ap}, & Z^{MC}_{e,p,ap}, & Z^{MC}_{o,p,ap}, \\
Z^{MC}_{e,ap,p}, & Z^{MC}_{o,ap,p}, & Z^{MC}_{e,p,p},  & Z^{MC}_{o,p,p}.
\end{array}
\end{equation}

If the top boundary is in the fermion condensed state given by $\tilde{v}_f$, the total partition function of this critical point is given by
\begin{equation}
Z^{MC}_f = Z^{MC}_{e,ap,ap} + Z^{MC}_{o,ap,ap} = Z^{MC}_{ap,ap,ap},
\label{eq:Z_fermion}
\end{equation}
where the first $a$ in the subscript of the final expression indicate normal boundary condition in the time direction (no symmetry operator inserted). This partition function is invariant under $S$, but only invariant under $T^2$ not $T$. This is expected for a system containing real fermions. 

If the top boundary is in the flux-loop condensed state given by $\tilde{v}_m$, the total partition function is given by
\begin{equation}
Z^{MC}_m = Z^{MC}_{e,ap,ap} + Z^{MC}_{e,ap,p} + Z^{MC}_{e,p,ap} + Z^{MC}_{e,p,p},
\label{eq:Z_f1form}
\end{equation}
which is a sum over all charge even sectors with different flux configurations. In the same way as was shown in~\cite{Hsieh2016}, one can find that this is a modular invariant partition function composed of the $2+1$D Majorana cone partition functions (also see Appendix \ref{app:majorana_cone}). It describes a bosonic critical theory with an emergent $\mathbb{Z}_2$ $1$-form symmetry.

If the top boundary is in the Majorana-chain-decorated-flux-loop condensed state given by $\tilde{v}_{mM}$, the total partition function is given by
\begin{equation}
Z^{MC}_{mM} = Z^{MC}_{e,ap,ap} + Z^{MC}_{e,ap,p} + Z^{MC}_{e,p,ap} + Z^{MC}_{o,p,p},
\end{equation}
which is a sum over three flux sectors with even charge and one flux sector $(p,p)$ with odd charge. Because with the $(p,p)$ boundadry condition the Majorana cone has a zero mode, the partition function with even charge $Z^{MC}_{e,p,p}$ is the same as the partition function with odd charge $Z^{MC}_{o,p,p}$. Therefore, with the top boundary in the $\tilde{v}_{mM}$ state, the sandwich still gives a modular invariant partition function of the $2+1$D Majorana cone.

\subsection{Phase diagrams}

\begin{figure}[ht]
    \centering
    \includegraphics[scale=0.75]{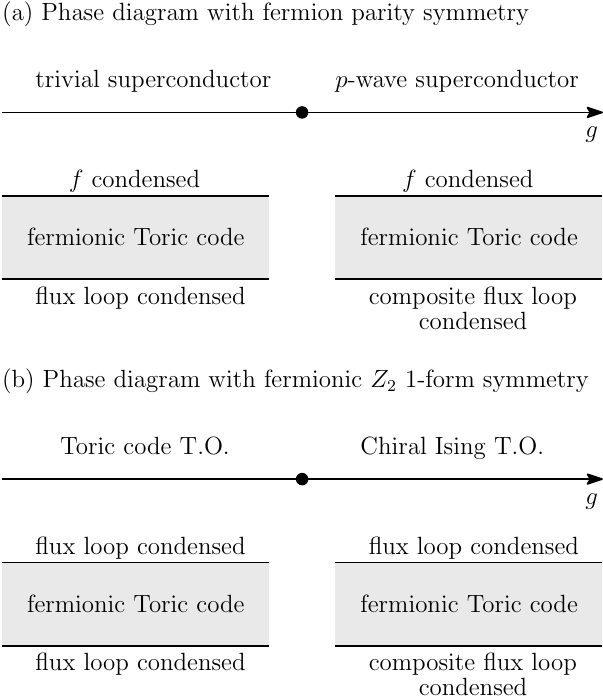}
    \caption{Phase diagrams of $2d$ systems and the ``sandwich constructions''. T.O. stands for topological order. The three types of gapped boundaries, $f$ condensed, flux loop condensed and the composite flux loop condensed boundary, have vectors of partition functions given by $\tilde{v}_f, \tilde{v}_m$ and $\tilde{v}_{mM}$, respectively. (a) The critical point is described by the theory with a single Majorana cone, whose partition function is given by (\ref{eq:Z_fermion}). (b) The low energy theory of the critical point is described by a bosonic theory with the partition function (\ref{eq:Z_f1form}). In fermionic toric code, the composite loop is a flux loop attached with a Majorana chain.}
    \label{fig:phase_diagram}
\end{figure}

The continuous transition between the trivial superconducting phase and the $p+ip$ superconducting phase is described by the critical Majorana cone theory with the Hamiltonian perturbed by a mass term $\delta H = -\frac{1}{(2\pi)^2}\int d^2\mathbf{r} m \lambda (\mathbf{r})^T\lambda (\mathbf{r})$. Indeed, this can be seen from the torus partition function. 
\begin{align}
    Z^{\text{Majorana}}_\mathbf{a}[-m]=Z^{\text{Majorana}}_\mathbf{a}[m]Z^{p+ip}_\mathbf{a},
\end{align}
where $Z^{p+ip}_\mathbf{a}$ is the partition function of the effective spin topological quantum field theory of the $p+ip$ superconducting state,\cite{gaiotto2016spin}
\begin{align}
    Z^{p+ip}_\mathbf{a}=\Tr_{a_x,a_y}e^{i 2\pi N_f(a_\tau-\frac{1}{2})},
\end{align}
with $N_f$ the total fermion number. Or rather
\begin{align}
    Z^{p+ip}_\text{p,p,p}=-1;~~\text{otherwise}~~~Z^{p+ip}_\mathbf{\mathbf{a}}=1.
    \label{eq:p+ipstacking}
\end{align}

This transition between two superconducting phases in a fermionic system helps us to learn about the transition between two topological ordered phases. We illustate the corresponding phase digrams in Figure \ref{fig:phase_diagram}.  Indeed, we can gauge the fermion parity symmetry in the fermionic system and obtain a bosonic system with emergent fermions. At low energy, the worldhistory of the emergent fermion generates the fermionic $\mathbb{Z}_2$ $1$-form symmetry. In terms of the sandwich construction, to gauge the fermion parity symmetry is to change the top boundary from the $f$ condensed one ($\tilde{v}_f$) to the flux loop condensed one ($\tilde{v}_m$). Correspondingly, the eight topological sectors inherited from the minimal entangled states are labeled in terms of symmetry charges and boundary conditions with respect to the $\mathbb{Z}_2$ $1$-form symmetry. Explicitly, $W_f^x=\pm 1$ ($W_f^y=\pm 1$) labels the $1$-form symmetry charge along the $x$ ($y$)direction. $V_m^{xy}=\pm 1$ labels whether there is a $1$-form symmetry twist defect in the bottom boundary, in other words whether there is a puncture where a fermion $f$ is inserted in the bottom boundary.

Upon gauging, the trivial superconducting phase becomes the phase with toric code topological order (TO); the $p$-wave superconducting phase becomes the phase with chiral Ising TO. In both TOs, the emergent fermionic $1$-form symmetry is spontaneously broken. 
The continuous transition point described by Eq (\ref{eq:Z_fermion}) becomes the continuous transition point between two TOs, with the bosonic partition function (\ref{eq:Z_f1form}). We have used the knowledge that in terms of partition functions, gauging a $0$-form symmetry is to sum over partition functions twisted by the symmetry defects along non-contractible cycles in the spacetime manifold. 

By virtue of the relation (\ref{eq:p+ipstacking}), we can further compare the spectrums on two sides of the transition point. The partition function $Z^{\mathcal{B}}$ near the bosonic phase transition can be expressed by the partition functions $Z^{\mathcal{F}}_{P^f,a_x,a_y}$ near the fermionic transition. When the mass is non-zero but small $|m|\gtrsim 0$,
\begin{itemize}
    \item On the toric code side, 
\begin{align}
Z^{\mathcal{B}}\left(|m|\right)=&Z^\mathcal{F}_{e,p,p}\left(|m|\right)+Z^\mathcal{F}_{e,ap,p}\left(|m|\right)\nonumber\\
&+Z^\mathcal{F}_{e,p,ap}\left(|m|\right)+Z^\mathcal{F}_{e,ap,ap}\left(|m|\right).
\end{align}
\item On the chiral Ising side,
\begin{align}
Z^{\mathcal{B}}\left(-|m|\right)=&Z^\mathcal{F}_{o,p,p}\left(|m|\right)+Z^\mathcal{F}_{e,ap,p}\left(|m|\right)\nonumber \\
&+Z^\mathcal{F}_{e,p,ap}\left(|m|\right)+Z^\mathcal{F}_{e,ap,ap}\left(|m|\right).
\end{align}
\end{itemize}

The spectrums on the two sides of the transition are very similar, except for the contribution from the ``p,p'' sectors. When the mass is small, each state in the spectrum $Z^{\mathcal{F}}_{e,p,p}(|m|)$ has a correspondence in the spectrum $Z^{\mathcal{F}}_{o,p,p}(|m|)$. The two differ only by a single point-like fermion excitation, and the energy difference is on the order of the mass $\delta E \sim |m|$. 
This is consistent with the physical picture that if on one side of the transition, the gapped ground state is a loop condensate, so is the ground state on the other side.

\subsection{Zero-point energy of the sectors}
\label{sec:energy}
In $1+1$D CFT, the energy and momentum of each topological sector are important quantities. Through the state-opertor correspondence, they are related to conformal dimensions of the primary fields. In this subsection, we compute the analogous quantities in 2+1D, the ``zero-point'' energy of different sectors of the 2+1D Majorana cone critical theory. We find that similar to $1+1$D CFTs, the energy difference between different sectors scales as $1/L$ at the $2+1$D critical point, where $L$ is the linear size of the system. When the system is gapped, the energy difference decays exponentially with $L$. This result is similar to that derived for Dirac fermions in Ref.~\onlinecite{Ishikawa2021}.

Our calculation is based on the following realization of the Majorana cone transition on a $2D$ square lattice of size $N_x\times N_y$
\begin{equation}
\begin{array}{ll}
H =  \sum_{m,n} & -t \left(c^{\dagger}_{m+1,n}c_{m,n} + c^{\dagger}_{m,n+1}c_{m,n} +h.c. \right)\\
& - (\mu-4t) c^{\dagger}_{m,n}c_{m,n} \\  & + \left(\Delta c^{\dagger}_{m+1,n}c^{\dagger}_{m,n} + i\Delta c^{\dagger}_{m,n+1}c^{\dagger}_{m,n}+h.c. \right),
\end{array}
\label{eq:2dH}
\end{equation}
where $c_{m,n}$ are complex fermion modes on lattice site $(m,n)$. At $\mu=0$, the system is gapless and realizes the Majorana cone critical point. It is the transition point between the trivial superconducting phase (at $\mu<0$) and the chiral $p+ip$ superconductor (at $0<\mu <4t$). 

In Appendix~\ref{app:energies}, we explain the details of the calculation. We find that, in the gapped phases ($\mu<0$ or $\mu>0$), the energy difference between the different flux sectors are exponentially small. Denote the zero-point energy with anti-periodic boundary condition in both $x$ and $y$ directions as $E_0^{ap,ap}$, the zero point energy with anti-periodic boundary condition in $x$ direction and periodic boundary condition in $y$ direction as $E_0^{ap,p}$ and the zero point energy with periodic boundary condition in both direction as $E_0^{p,p}$, we find
\begin{equation}
E_0^{ap,ap}-E_0^{ap,p} \sim e^{-\alpha L}, E_0^{ap,p}-E_0^{p,p} \sim e^{-\alpha' L},
\end{equation}
as shown in Fig.~\ref{fig:2DdEgap}. This is consistent with the fact that once these gapped superconducting states are coupled to dynamical $\mathbb{Z}_2$ gauge fields, they become topological states with degenerate ground state. The lowest-energy states from different flux sectors form the degenerate ground space. In the trivial phase (with $\mu<0$), the lowest-energy states in the four sectors all have even fermion parity and become the four-fold degenerate ground space once gauged. For the $p+ip$ phase (with $\mu>0$) on the other hand, the lowest energy states in three sectors have even fermion parity, while the one in the last sector (the $(p,p)$ sector) has odd fermion parity. Therefore, after coupling to dynamical gauge field, there is a three-fold degeneracy. This is consistent with the degeneracy computed from the inner product $\langle \tilde v_m |\tilde v_{mM}\rangle =3$.

At the critical point ($\mu=0$), the energy difference between the flux sectors scales as $1/L$. 
\begin{equation}
E_0^{ap,ap}-E_0^{ap,p} \sim \frac{1}{L},~~ E_0^{ap,p}-E_0^{p,p} \sim \frac{1}{L},
\end{equation}
as shown in Fig.~\ref{fig:2DdEgapless}.

In Table~\ref{table:scaling}, we list the normalized energy difference with normalization factor $2\Big / \epsilon\left(\frac{4\pi}{N_x},0\right)$. $\epsilon\left(\frac{4\pi}{N_x},0\right)$ is the energy of the lowest excited state with momentum $\left(\frac{4\pi}{N_x},0\right)$ in the $(e,ap,ap)$ sector. At a Majorana cone, $\epsilon\left(\frac{4\pi}{N_x},0\right)$ scales as $1/N_x$, therefore, the normalized energy difference converge to a constant (listed in the table). As shown in the table, the constant depends on the aspect ratio between $N_x$ and $N_y$. If we keep $N_y$ fixed while increasing $N_x$ (such that $N_x/N_y\to \infty$), we approach the limit of a $1+1$D chain in the $x$ direction. We see that in this limit, the normalized energy difference between the $(ap,p)$ and $(p,p)$ sectors is $1/8$, which is exactly the expected total conformal dimension of the flux primary field in the $1+1$D gapless Majorana chain ($\sigma\bar{\sigma}$ in table~\ref{table:1DIsing}). In the opposite limit of $N_x/N_y \to 0$, we get a $1+1$D chain in the $y$ direction. In this limit, the $(ap,ap)$ and $(ap,p)$ sectors correspond to the two flux sectors of a massive $1+1$D chain. Their energy difference goes to zero, also expected. Ref.~\onlinecite{Luo2023} studied the dependence of the zero-point energy (Casimir energy) of $2+1$D bosonic CFTs on the modular parameters of the torus. A generalized version of that discussion may apply to the fermion zero-point energy discussed here.

\begin{table}[th]
    \centering
    \begin{tabular}{||c|c|c||}
        $N_x/N_y$ & $(ap,ap)-(ap,p)$ & $(ap,p)-(p,p)$ \\[2pt]
         $\infty$ &  & $0.1250$\\
         $3$ & $0.4379$ & $0.1250$\\
         $2$ & $0.1721$ & $0.1250$ \\
         $1.5$ & $0.0808$ & $0.1251$ \\
         $1$ & $0.0217$ & $0.1262$\\
         $5/6$ & $0.0104$ & $0.1283$ \\
         $5/7$ & $0.0051$ & $0.1316$ \\
         $2/3$ & $0.0036$ & $0.1337$ \\
         $1/2$ & $0.0006$ & $0.1478$ \\
         $1/3$ & $0.0000$ & $0.1876$ \\
         $1/4$ & $0.0000$ & $0.2339$ \\
    \end{tabular}
    \caption{Left column: aspect ratio of the 2D lattice; Middle column: normalized energy difference between the $(ap,ap)$ and $(ap,p)$ sector; Right column: normalized energy difference between the $(ap,p)$ and $(p,p)$ sector. }
    \label{table:scaling}
\end{table}

At the critical point ($\mu=0$), the sectors induced by inserting $mM$ flux defects ($\pi$ flux decorated with a Majorana chain) have exactly the same energy as those induced by the bare $m$ flux. This is because the only difference between these two types of sectors is when defects are inserted in both $x$ and $y$ directions, the $(p,p)$ sector induced by $m$ flux has even fermion parity while the $(p,p)$ sector induced by $mM$ flux has odd fermion parity. But at the critical point the $(p,p)$ sector has a zero mode. Therefore, the lowest energy state with fermion parity even has exactly the same energy as that with fermion parity odd.

From this calculation, we see how the Majorana cone transition embodies the competition between two loop condensates in quantitative ways. On the two sides of the transition, flux loops are condensed. Therefore, the zero energy difference between the different flux sectors becomes very small (exponentially small). The two gapped phases correspond to two different loop condensates: the trivial phase condenses with `bare' flux loop while the $p+ip$ phase condenses the flux loop dressed with the Majorana chain. This translates into the difference in the fermion parity of the lowest energy state in the four flux sectors, with the $(p,p)$ sector of the $p+ip$ state being fermion odd, while all other sectors are fermion even. At the transition, the two condensates compete with each other, so that neither of the flux loops can have an exponentially small energy. Instead, they acquire a polynomially small energy ($\sim 1/L$). Their composite, the Majorana chain, also has a $\sim 1/L$ energy. This is because adding a Majorana chain around a nontrivial cycle can be achieved with a unitary transformation (a sequential quantum circuit\cite{Chen2024,Tantivasadakarn2024}) except at the last step when the nontrivial cycle closes. At this step, an extra fermion may need to be added depending on the flux through the cycle. Therefore, the energy cost of adding a Majorana chain to a particular sector is at most the cost of a single fermion, which scales as $\sim 1/L$ at the critical point.

Note that in order to do this calculation, we need to put the Hamiltonian on a 2d torus and insert flux along the nontrivial cycles. If the flux defect goes along a trivial cycle, the energy difference would be zero because the Hamiltonian with and without the defect are related unitarily by applying the fermion parity $\mathbb{Z}_2$ symmetry on the area inside of the trivial cycle. Therefore, there is no analogous calculation on a 2d sphere. On the other hand, the 2d torus does not have conformal symmetry and loses many features of conformal field theory including operator state correspondence\cite{Belin2018}.

\section{$2+1$D spin system with $\mathbb{Z}_2$ symmetry}
\label{sec:Z2}

A $2+1$D spin system with 0-form $\mathbb{Z}_2$ symmetry can be realized on the boundary of the $3+1$D bosonic Toric Code ($3+1$D $\mathbb{Z}_2$ gauge theory with bosonic charge). The sandwich structure is very similar to that shown in Fig.~\ref{fig:fZ2}. In this section, we discuss the bulk properties of the $3+1$D bosonic Toric Code, its gapped boundaries and the implication of its line excitations on the topological sectors on the boundary theory. We leave the study of specific boundary critical theories to the future.

\subsection{$3+1$D bosonic Toric Code: bulk}

The $3+1$D Toric Code has an eight-dimensional ground space on 3D torus. Similar to the fermionic Toric Code case, we can choose a basis for the ground space such that the gauge charge string operators and the gauge flux membrane operators act as logical $X$ and $Z$'s on the three logical qubits
\begin{equation}
\begin{array}{lll}
W^x_e \sim Z_1, & W^y_e \sim Z_2, & W^z_e \sim Z_3, \\
V^{yz}_{\phi} \sim X_1, & V^{zx}_{\phi} \sim X_2, & V^{xy}_{\phi} \sim X_3. \\
\end{array}
\label{eq:TC_WV}
\end{equation}
Here we use $e$ to denote gauge charge and $\phi$ to denote $\pi$ flux to distinguish from the fermionic Toric Code case.

The descendent line excitation in the $3+1$D toric code is the ``Cheshire string'', along which gauge charge $e$ condensates. The Cheshire string is different from the Majorana chain excitation in that it is non-invertible. The fusion of two Cheshire strings, denoted as $C$, results in a single Cheshire string with a coefficient.
\begin{equation}
C \times C  = \mathcal{Z}_2 \ C,
\end{equation}
The coefficient comes from a decoupled $1+1$D $\mathbb{Z}_2$ gauge theory. 

If we use a membrane operator $\mathcal{V}^{yz}_C$ to create a pair of Cheshire strings along the $y$ direction and then pull them apart in the $z$ direction \cite{tantivasadakarn2024string}, what action does it have on the ground space? We find that the induced operation is a quantum channel instead of a simple unitary operator. We thus use $\mathcal{V}$ to denote the membrane operator instead of $V$.

First, Cheshire strings along the $y$ direction implement a projection onto one of the eigenspaces of $W^y_e$. As the Cheshire string is a charge condensate, charges hopping along the string should acquire the same phase factor once they are back to the original point. This phase factor is either $1$ or $-1$, which corresponds exactly to the eigenvalue of $W^y_e$. Secondly, with a pair of Cheshire strings, there is a two-fold degeneracy corresponding to each of the Cheshire strings carrying even or odd total charge. These two cases can be mapped into each other by tunneling a pair of charges in between. As the two Cheshire strings are pulled apart in the $z$ direction, any tunneled charges are pulled apart at the same time, implementing or not implementing the string operator $W^z_e$ in the process. Therefore, on the three qubit logical space where the charge string and flux membrane operators act as in Eq.~\ref{eq:TC_WV}, the logical operation induced by $\mathcal{V}^{yz}_C$ is a quantum channel $\rho\rightarrow S(\rho), S(\rho)=\sum_{k}M_k \rho M_k^\dagger$ with Kraus operators $M_k$ given by, (up to a normalization factor $\frac{1}{2}$)
\begin{equation}
\mathcal{V}^{yz}_C: |0\rangle \langle 0|_2, \ |1\rangle \langle 1|_2, Z_3|0\rangle \langle 0|_2, \ Z_3|1\rangle \langle 1|_2,
\end{equation}
where the subscript denotes the logical qubit.

Or equivalently, 
\begin{equation}
\mathcal{V}^{yz}_C: |00\rangle \langle 00|_{23}, \ |01\rangle \langle 01|_{23}, |10\rangle \langle 10|_{23}, \ |11\rangle \langle 11|_{23}.
\label{eq:VC}
\end{equation}
We give an explicit lattice derivation of this action in Appendix~\ref{sec:VC}.

The eight-fold degenerate ground space transforms unitarily under the modular transformations of the 3D torus. It takes very similar form to the modular transformation for the fermionic Toric Code in Eq.~\ref{eq:fTC_MT}, with slight but important difference.
\begin{equation}
S = \begin{pmatrix} 1 & 0 & 0 & 0 & 0 & 0 & 0 & 0 \\ 0 & 0 & 0 & 0 & 1 & 0 & 0 & 0\\ 0 & 1 & 0 & 0 & 0 & 0 & 0 & 0 \\ 0 & 0 & 0 & 0 & 0 & 1 & 0 & 0 \\ 0 & 0 & 1 & 0 & 0 & 0 & 0 & 0 \\ 0 & 0 & 0 & 0 & 0 & 0 & 1 & 0 \\ 0 & 0 & 0 & 1 & 0 & 0 & 0 & 0 \\ 0 & 0 & 0 & 0 & 0 & 0 & 0 & 1 \end{pmatrix}, 
T = \begin{pmatrix} 1 & 0 & 0 & 0 & 0 & 0 & 0 & 0 \\ 0 & 1 & 0 & 0 & 0 & 0 & 0 & 0\\ 0 & 0 & 0 & 1 & 0 & 0 & 0 & 0 \\ 0 & 0 & 1 & 0 & 0 & 0 & 0 & 0 \\ 0 & 0 & 0 & 0 & 1 & 0 & 0 & 0 \\ 0 & 0 & 0 & 0 & 0 & 1 & 0 & 0 \\ 0 & 0 & 0 & 0 & 0 & 0 & 0 & 1 \\ 0 & 0 & 0 & 0 & 0 & 0 & 1 & 0 \end{pmatrix}. 
\label{eq:TC_MT}
\end{equation}
The derivation of the $S$ and $T$ matrices can again be found in Appendix~\ref{sec:WW}.

There are two modular invariant (eigenvalue-1) vectors under both $S$ and $T$
\begin{align}
u_{e} = \begin{pmatrix} 1 & 0 & 0 & 0 & 0 & 0 & 0 & 0\end{pmatrix}, \\
u_{\phi} = \begin{pmatrix} 1 & 1 & 1 & 1 & 1 & 1 & 1& 1 \end{pmatrix}.
\end{align}
Note that $u_e$ is the common eigenvalue-1 eigenstate of $W^x_e$, $W^y_e$, $W^z_e$ while $u_{\phi}$ is the common eigenvalue-1 eigenstate of $V^{yz}_{\phi}$, $V^{zx}_{\phi}$, $V^{xy}_{\phi}$, hence the labeling. Moreover, we will see that $u_e$ and $u_{\phi}$ correspond to the charge condensed and flux loop condensed boundaries of the $3+1$D bosonic Toric Code. 

\subsection{$3+1$D bosonic Toric Code: gapped boundaries}

To see the correspondence of $u_e$ and $u_{\phi}$ to gapped boundaries, we map to the Minimally Entangled State basis which are common eigenvectors of $W^x_e$, $W^y_e$, and $V^{xy}_{\phi}$. The vectors map to
\begin{align}
\tilde{u}_{e} = \begin{pmatrix} 1 & 1 & 0 & 0 & 0 & 0 & 0 & 0\end{pmatrix}, \\
\tilde{u}_{\phi} = \begin{pmatrix} 1 & 0 & 1 & 0 & 1 & 0 & 1& 0 \end{pmatrix}.
\end{align}

Then we physically cut the $3+1$D system open across the $xy$ plane. The eight ground states on the 3D torus now become the eight topological sectors of the system with open boundaries labeled by the eigenvalues of $W^x_e$, $W^y_e$, and $V^{xy}_{\phi}$ that indicate the anti-periodic / periodic
boundary condition in the $x$ direction, anti-periodic / pe-
riodic boundary condition in the $y$ direction, and even /
odd total charge on each boundary. 

With this interpretation, we see that $\tilde{u}_e$ corresponds to the superposition of the no-flux-even-charge sector and the no-flux-odd-charge sector. Therefore, it describes the boundary state of bosonic charge condensate. The $\tilde{u}_{\phi}$ vector corresponds to the superposition of the no-flux-even-charge sector, the $x$-flux-even-charge sector, the $y$-flux-even-charge sector and the $x$-flux-$y$-flux-even-charge sector. Therefore,
it describes the boundary state of the flux loop condensate.

Note that again not all gapped boundary types of the $3+1$D bosonic Toric Code are captured using this method. There is another loop-condensed boundary that correspond to pumping a $2+1$D $\mathbb{Z}_2$ SPT to the loop-condensed boundary described by $\tilde{u}_{\phi}$. The method we discuss here does not capture this one (or does not distinguish between the two). 

\subsection{$2+1$D Ising theory}

With the understanding of gapped boundaries of $3+1$D bosonic Toric Code, we can now build the $2+1$D sandwich for the $2+1$D bosonic system with $\mathbb{Z}_2$ symmetry (coupled or not coupled to the dynamical $\mathbb{Z}_2$ gauge field).

Similar to the fermionic Toric Code case, if the top boundary of the sandwich is gapped by the condensation of charges, the relevant symmetry of the sandwich is a $0$-form $\mathbb{Z}_2$ symmetry given by $V^{xy}_{\phi}$. If the top boundary is gapped by condensing flux loops, the relevant symmetry of the sandwich is a $1$-form $\mathbb{Z}_2$ symmetry generated by the string operator $W_e$ in the $xy$ plane. Table~\ref{table:Z2_gapped} lists the gapped phases realized when both boundaries are gapped. 

\begin{table}[th]
    \centering
    \begin{tabular}{||c|c|c||}
        Top & Bottom & Gapped Phase \\[2pt]
         $\tilde{u}_e$ & $\tilde{u}_e$ & $\mathbb{Z}_2$ symmetry breaking\\
         $\tilde{u}_e$ & $\tilde{u}_{\phi}$ & $\mathbb{Z}_2$ symmetric\\
         $\tilde{u}_e$ & $\tilde{u}'_{\phi}$ & $\mathbb{Z}_2$ SPT \\
         $\tilde{u}_{\phi}$ & $\tilde{u}_{e}$ & trivial $1$-form $\mathbb{Z}_2$ symmetric \\
         $\tilde{u}_{\phi}$ & $\tilde{u}_{\phi}$ & $2+1$ Toric Code \\
         $\tilde{u}_{\phi}$ &
         $\tilde{u}'_{\phi}$ & $2+1$ double semion topo order \\[1pt]
    \end{tabular}
    \caption{Gapped phases in the $2+1$D sandwich structure with $3+1$D bosonic Toric Code in the bulk. The top and bottom boundaries are gapped by condensing the bosonic gauge charge ($\tilde{u}_e$) or condensing the flux loop ($\tilde{u}_{\phi}$ and $\tilde{u}'_{\phi}$). We use $\tilde{u}_{\phi}$ and $\tilde{u}'_{\phi}$ to denote the two different flux loop condensing boundaries although we only find one modular invariant vector to describe them.}
    \label{table:Z2_gapped}
\end{table}

Therefore, with the bosonic Toric Code sandwich, we can study transitions between $2+1$D $\mathbb{Z}_2$ symmetric phase and symmetry breaking phase, $2+1$D trivial and nontrivial SPTs with $\mathbb{Z}_2$ symmetry, as well as the gauged version (coupled to dynamical $\mathbb{Z}_2$ gauge field) of these transitions. Ref.~\cite{Chen2016} discussed a gapless boundary state of the $3+1$D bosonic Toric Code whose partition function sectors transform covariantly under the bulk modular transformation given in Eq.~\ref{eq:TC_MT}. We leave more careful analysis of the critical points between different gapped boundary state to future work, but we do want to comment on the effect the Cheshire string excitation has on the boundary topological sector. 

Suppose that we start from a particular sector of the sandwich labeled by the eigenvalue of $W^x_e$, $W^y_e$ and $V^{xy}_{\phi}$. Tunneling a Cheshire string along the $y$ direction in the $z$ direction projects into the eigensector of $W^y_e$. If we already started in such a sector, this part of the mapping is trivial. Secondly, the Cheshire string will tunnel or not tunnel a charge in the $z$ direction as a quantum channel. Therefore, it maps a single eigensector under $V^{xy}_{\phi}$ into a direct sum of two sectors. This, of course is a direct consequence of the degeneracy associated with the Cheshire string. This will be potentially relevant when we study sandwich structures with a $\tilde{u}_e$ top boundary where charges condense, and hence Cheshire strings condense\cite{Kong2024higher}.

\section{$2+1$D spin system with anomalous $\mathbb{Z}_2\times \mathbb{Z}_2$ symmetry}
\label{sec:Z2Z2}

In the previous two cases, when we discuss the degenerate ground states of the bulk topological order or the topological sectors in the sandwich structure, we did not have to involve descendent excitations / defects because we can reach all ground states / topological sectors by tunneling elementary excitations / defects. However, when the bulk topological order is the $3+1$D $\mathbb{Z}_2\times \mathbb{Z}_2$ twisted gauge theory characterized by the nontrivial cocycle in $H^4(\mathbb{Z}_2\times \mathbb{Z}_2,U(1))$ and the sandwich is a $2+1$D spin system with anomalous $\mathbb{Z}_2\times \mathbb{Z}_2$ 0-form symmetry, we have to include them in the discussion. This is because the flux loop excitations in the $3+1$D $\mathbb{Z}_2\times \mathbb{Z}_2$ gauge theory have to be ``Cheshire''. That is, for the $m_1$ flux loop to be gapped, $e_2$ has to condense on it. Similarly, for the $m_2$ flux loop to be gapped, $e_1$ has to condense on it. When two $m_1$ flux loops fuse, the fusion result is a $C_2$, a Cheshire string of charge $e_2$. Similary, when two $m_2$ flux loops fuse, the fusion result is a $C_1$, a Cheshire string of charge $e_1$.

We can still use $\mathcal{V}^{xy}_{\phi^I}$, $W^{x}_{e^I}$, $W^{y}_{e^I}$, $\mathcal{V}^{xy}_{\phi^{II}}$, $W^{x}_{e^{II}}$ and $W^{y}_{e^{II}}$ to label the Minimally Entangled States or the topological sectors even though the gauge flux membrane operators are not simple unitaries any more. The gauge charge string operators $W^{x}_{e^I}$, $W^{y}_{e^I}$, $W^{x}_{e^{II}}$ and $W^{y}_{e^{II}}$ are simple $\mathbb{Z}_2$ unitaries so we can use their $\pm 1$ eigenvalues to label the sectors. Once the eigenvalues of $W^{x}_{e^k}$, $W^{y}_{e^k}$ are fixed, the MES / sector is invariant under the $xy$ plane Cheshire membrane operator $\mathcal{V}_{C^k}$, $k=I,II$. Therefore, all six operators share a common set of fixed point MES / sectors. The eigenvalues of $W^{x}_{e^I}$, $W^{y}_{e^I}$, $W^{x}_{e^{II}}$ and $W^{y}_{e^{II}}$ are good quantum numbers labeling the MES / sectors. $\mathcal{V}^{xy}_{\phi^I}$ and $\mathcal{V}^{xy}_{\phi^{II}}$ do not give distinct eigenvalues for the fixed point sectors. We are just going to label the MES / sectors distinguished due to $\mathcal{V}^{xy}_{\phi^k}$ as $a^k$ and $b^k$. The sectors are hence labeled by $(\mu^1,m^1, n^1, \mu^2, m^2, n^2)$ where $\mu \in \{a,b\}$, $m,n=\pm 1$.

Following the previous discussion, we can write down the action on the topological sectors by different line defects. A Cheshire line defect generated by $\mathcal{V}^{yz}_{C^I}$ maps the sectors as
\begin{equation}
\begin{array}{ll}
\mathcal{V}^{yz}_{C^I}: &(\mu^1,m^1, n^1, \mu^2, m^2, n^2) \to \\
 & (\mu^1,m^1, n^1, \mu^2, m^2, n^2)\oplus (\bar{\mu}^1,m^1, n^1, \mu^2, m^2, n^2)
\end{array}
\end{equation}
A flux line defect generated by $\mathcal{V}^{yz}_{\phi^I}$ maps the sectors as
\begin{equation}
\begin{array}{ll}
\mathcal{V}^{yz}_{\phi^I}: &(\mu^1,m^1, n^1, \mu^2, m^2, n^2) \to \\
 & (\mu^1,\bar{m}^1, n^1, \mu^2, m^2, n^2)\oplus (\mu^1,\bar{m}^1, n^1, \bar{\mu}^2, m^2, n^2)
\end{array}
\end{equation}
The point defect generated by $W^{z}_{e^I}$ maps the sectors as
\begin{equation}
\begin{array}{ll}
\mathcal{W}^{z}_{e^I}: &(\mu^1,m^1, n^1, \mu^2, m^2, n^2) \to \\
 & (\bar{\mu}^1,m^1, n^1, \mu^2, m^2, n^2)
\end{array}
\end{equation}

Apart from the Cheshire strings, the $\mathbb{Z}_2\times \mathbb{Z}_2$ gauge theory also has an invertible descendent line excitation corresponding to a $1+1$D SPT state with $\mathbb{Z}_2\times \mathbb{Z}_2$ symmetry. The key feature of the $1+1$D SPT is that the flux of the first $\mathbb{Z}_2$ is tied to the charge of the second $\mathbb{Z}_2$ and vice versa. Therefore, the unitary $\mathbb{Z}_2$ symmetry associated with sweeping this invertible defect (for example, one along $y$ direction) implements two controlled-NOT operations one with $n^1$ as the control, $\mu^2$ as the target and the other with $n^2$ as the control, $\mu^1$ as the target.
\begin{equation}
\begin{array}{ll}
V^{yz}_I: &(\mu^1,m^1, n^1, \mu^2, m^2, n^2) \to \\
 & ((\mu^1)^{n^2},m^1, n^1, (\mu^2)^{n^1}, m^2, n^2),
\end{array}
\end{equation}
where $(\mu^k)^{-1}\doteq \bar{\mu}^k$. We can explicitly check that the mapping induced by the defects is consistent with their fusion rule. 

Although the bulk has $\mathbb{Z}_2\times \mathbb{Z}_2$ topological order, this sandwich structure is relevant for the transition between the $2+1$ D trivial and non-trivial SPTs with $\mathbb{Z}_2$ symmetry. The $2+1$D SPTs can first be realized in a sandwich structure with the $3+1$D bosonic Toric Code in the bulk, the gauge charge condensed at the top boundary, and the gauge flux condensed at the bottom boundary, as discussed in the previous section. Two different ways to condense gauge flux on the bottom boundary lead to two different SPTs. A bulk $\mathbb{Z}_2$ symmetry keeps the bulk invariant while exchanging the two flux-condensed boundaries.\cite{ji2023boundary} Gauging this bulk $\mathbb{Z}_2$ symmetry leads to a sandwich structure with the $\mathbb{Z}_2\times \mathbb{Z}_2$ twisted topological order discussed above. Therefore, if neither $\mathbb{Z}_2$ symmetry breaks at the bottom boundary (neither charge condenses), the sandwich is at the critical point transitioning between the $\mathbb{Z}_2$ SPTs. On the other hand, due to the twisted nature of the $\mathbb{Z}_2\times \mathbb{Z}_2$ topological order in the bulk, the boundary can have an anomalous $\mathbb{Z}_2\times \mathbb{Z}_2$ symmetry and there can be a deconfined critical point between different $\mathbb{Z}_2$ symmetry breaking phases. Ref.~\cite{Chen2016} also discussed a gapless boundary state of the $3+1$D twisted $\mathbb{Z}_2\times \mathbb{Z}_2$ gauge theory whose partition function sectors transform covariantly under the bulk modular transformation. We leave the study of specific critical points in this system to the future.


\section{Discussion}

In this paper, we discuss how line excitations in $3+1$D topological states become line defects in the $2+1$D (potentially gapless / critical) theories through the topological holography/symmetry TFT formalism. We focus on descendent line excitations in the bulk and studied their effect on the topological sectors of the boundary theory. 

Our analysis is most complete in the case of the $2+1$D superconducting system (with or without coupling to the dynamical $\mathbb{Z}_2$ gauge field) as a sandwich with the $3+1$D fermionic $\mathbb{Z}_2$ gauge theory in the bulk. The descendent Majorana chain excitation in the bulk corresponds to a unitary controlled-NOT transformation in the bulk ground space. Because of its existence, we find two distinct types of flux loop-condensed gapped boundaries corresponding to the $2+1$D trivial and $p+ip$ superconducting states respectively once put into the sandwich structure. The transition between the two superconducting states is known to be the critical Majorana cone state. We recover the modular invariant partition function of the Majorana cone by matching its topological sectors with those given by the loop-condensed gapped states. We further calculated the energy difference between different flux sectors and found similar scaling as in $1+1$D critical theories. 

We also explored $2+1$D spin systems with $\mathbb{Z}_2$ symmetry / anomalous $\mathbb{Z}_2\times \mathbb{Z}_2$ symmetry as sandwich structures with the $3+1$D bosonic $\mathbb{Z}_2$ gauge theory / twisted $\mathbb{Z}_2\times \mathbb{Z}_2$ gauge theory in the bulk. Here, the focus is on the non-invertible Cheshire string excitations in the bulk and discussed how they correspond to projection operations (or quantum channels) in the bulk ground space. On the boundary, the Cheshire line defects project onto the corresponding combination of the topological sectors. These defects will play a role in understanding the structure of the transition points between $2+1$D SPTs with $\mathbb{Z}_2$ symmetry as well as the deconfined critical point between two partial symmetry breaking phases under the $\mathbb{Z}_2\times \mathbb{Z}_2$ anomalous symmetry. We leave this part to future study.

If our understanding about the relation between $2+1$D topological order and $1+1$D CFT is a guide, there is much more work to do in one higher dimension, and what we discuss here is just the starting point. For example, in $1+1$D CFT, the conformal dimensions of the primary fields play a major role in constraining the possible structure of the CFT. We expect the scaling coefficient listed Table~\ref{table:scaling} to also play an important role in the structure of $2+1$D critical theories. The complete theory to describe bulk excitations of $3+1$D topological state using 2-category theory is still being developed. Everything we learn about the 2-category structure of the bulk excitations in $3+1$D will then inform us about the structure of line-defects and topological sectors in $2+1$D theories at their boundaries. 



\begin{acknowledgments}
We are indebted to inspiring discussions with Robijn Vanhove, Laurens Lootens, Xiao-Gang Wen, Shinsei Ryu, Alexei Kitaev, David Simmons-Duffin and Max Metlitski. X.C. is supported by the Walter Burke Institute for Theoretical Physics at Caltech, the Simons Investigator Award (award ID 828078) and the Institute for Quantum Information and Matter at Caltech. W.J. and X.C. are supported by the Simons collaboration on ``Ultra-Quantum Matter'' (grant number 651438). 
\end{acknowledgments}

\bibliography{references}

\appendix

\section{Walker Wang realization of Bosonic / Fermionic Toric Code}
\label{sec:WW}

The $3+1$D bosonic / fermionic Toric Code has a simple realization as the Walker-Wang model with a single $\mathbb{Z}_2$ boson / fermion as the input. 

\begin{figure}[ht]
    \centering
    \includegraphics[scale=0.60]{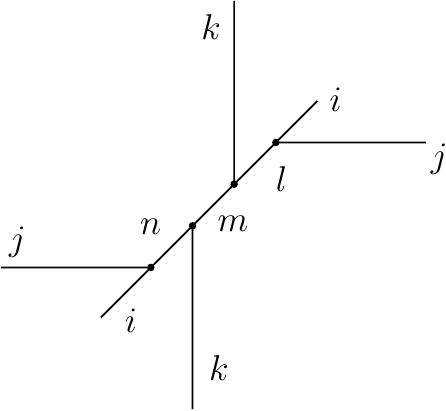}
    \caption{A minimal 3D Walker-Wang lattice with periodic boundary conditions.}
    \label{fig:WW}
\end{figure}

Following the discussion in Ref.~\cite{Wang2017}, consider the Walker-Wang model on a minimal 3D lattice with periodic boundary condition, as shown in Fig.~\ref{fig:WW}. All labels take value in $0$ and $1$. The two $i$ edges are periodically connected, so are the two $j$ edges and the two $k$ edges. When $l$, $m$, and $n$ satisfy $l = n = i+ j \text{\ mod\ } 2$, $m = i+ j +k \text{\ mod\ } 2$, the configuration is in the ground space. The eight configurations with $i,j,k = 0,1$ correspond exactly to the eight degenerate ground states on the 3D torus. 

For the bosonic Toric Code, the membrane and string operators act on the $\{ijk\}$ basis states as
\begin{equation}
\begin{array}{lll}
W^x_e \sim X_1, & W^y_e \sim X_2, & W^z_e \sim X_3 \\
V^{yz}_{\phi} \sim Z_1, & V^{zx}_{\phi} \sim Z_2, & V^{xy}_{\phi} \sim Z_3 \\
\end{array}
\label{eq:TC_WV2}
\end{equation}
In this basis, the modular transformation matrices are given by
\begin{equation}
S = \begin{pmatrix} 1 & 0 & 0 & 0 & 0 & 0 & 0 & 0 \\ 0 & 0 & 0 & 0 & 1 & 0 & 0 & 0\\ 0 & 1 & 0 & 0 & 0 & 0 & 0 & 0 \\ 0 & 0 & 0 & 0 & 0 & 1 & 0 & 0 \\ 0 & 0 & 1 & 0 & 0 & 0 & 0 & 0 \\ 0 & 0 & 0 & 0 & 0 & 0 & 1 & 0 \\ 0 & 0 & 0 & 1 & 0 & 0 & 0 & 0 \\ 0 & 0 & 0 & 0 & 0 & 0 & 0 & 1 \end{pmatrix}, 
T = \begin{pmatrix} 1 & 0 & 0 & 0 & 0 & 0 & 0 & 0 \\ 0 & 0 & 0 & 1 & 0 & 0 & 0 & 0\\ 0 & 0 & 1 & 0 & 0 & 0 & 0 & 0 \\ 0 & 1 & 0 & 0 & 0 & 0 & 0 & 0 \\ 0 & 0 & 0 & 0 & 1 & 0 & 0 & 0 \\ 0 & 0 & 0 & 0 & 0 & 0 & 0 & 1 \\ 0 & 0 & 0 & 0 & 0 & 0 & 1 & 0 \\ 0 & 0 & 0 & 0 & 0 & 1 & 0 & 0 \end{pmatrix} 
\label{eq:TC_MT2}
\end{equation}
To map to the basis where the logical operators take the form in Eq.~\ref{eq:TC_WV}, we need to apply the Hadamard transformation to all three logical qubits $H\otimes H \otimes H$, with $H = \frac{1}{\sqrt{2}} \begin{pmatrix} 1 & 1 \\ 1 & -1\end{pmatrix}$. We can check that after the basis transformation, the $S$ and $T$ matrices are mapped exactly to the form in Eq.~\ref{eq:TC_MT}.

For the fermionic Toric Code, the membrane and string operators act on the $\{ijk\}$ basis states as
\begin{equation}
\begin{array}{lll}
W^x_f \sim X_1Z_2Z_3, & W^y_f \sim X_2Z_1Z_3, & W^z_f \sim X_3Z_1Z_2 \\
V^{yz}_m \sim Z_1, & V^{zx}_m \sim Z_2, & V^{xy}_m \sim Z_3 \\
\end{array}
\label{eq:fTC_WV2}
\end{equation}
In this basis, the modular transformation matrices are given by
\begin{align}
S =& \begin{pmatrix} 1 & 0 & 0 & 0 & 0 & 0 & 0 & 0 \\ 0 & 0 & 0 & 0 & 1 & 0 & 0 & 0\\ 0 & 1 & 0 & 0 & 0 & 0 & 0 & 0 \\ 0 & 0 & 0 & 0 & 0 & 1 & 0 & 0 \\ 0 & 0 & 1 & 0 & 0 & 0 & 0 & 0 \\ 0 & 0 & 0 & 0 & 0 & 0 & 1 & 0 \\ 0 & 0 & 0 & 1 & 0 & 0 & 0 & 0 \\ 0 & 0 & 0 & 0 & 0 & 0 & 0 & 1 \end{pmatrix},\nonumber\\ 
T =& \begin{pmatrix} 1 & 0 & 0 & 0 & 0 & 0 & 0 & 0 \\ 0 & 0 & 0 & 1 & 0 & 0 & 0 & 0\\ 0 & 0 & 1 & 0 & 0 & 0 & 0 & 0 \\ 0 & 1 & 0 & 0 & 0 & 0 & 0 & 0 \\ 0 & 0 & 0 & 0 & 1 & 0 & 0 & 0 \\ 0 & 0 & 0 & 0 & 0 & 0 & 0 & -1 \\ 0 & 0 & 0 & 0 & 0 & 0 & 1 & 0 \\ 0 & 0 & 0 & 0 & 0 & -1 & 0 & 0 \end{pmatrix}. 
\label{eq:fTC_MT2}
\end{align}
To map to the basis where the logical operators take the form in Eq.~\ref{eq:fTC_WV}, we need to first apply controlled-$Z$ operation between qubit 1 and 2, qubit 2 and 3, qubit 3 and 1 before applying the Hadamard transformation to all three logical qubits $H\otimes H \otimes H$, with $H = \frac{1}{\sqrt{2}} \begin{pmatrix} 1 & 1 \\ 1 & -1\end{pmatrix}$. We can check that after this set of basis transformations, the $S$ and $T$ matrices are mapped exactly to the form in Eq.~\ref{eq:fTC_MT}.

\section{Partition functions of the theory of a single Majorana cone in $2+1$d}\label{app:majorana_cone}
On a $T^3$ torus, the Majorana cone partition function depends on the boundary conditions along the three directions, which can be either anti-periodic (AP) or periodic (P). Thus, there are $8$ components of partition functions, 
\begin{align}    Z^{\text{Majorana}}_{a_\tau,a_x,a_y}(g).
\end{align}
where $a_\tau,a_x,a_y$ take values in $0$ (P) or $\frac{1}{2}$ (AP), and label the boundary condition along the $\tau,x$ and $y$ direction, respectively. And $g$ stands for the Euclidean metric, which can be expressed in terms of modular parameters of a flat three-torus.\cite{Hsieh2016}

The explicit partition function can be computed following the procedure in \cite{Hsieh2016}, illustrated for the Dirac fermion. When the fermion field has periodic boundary conditions along all three directions on a $T^3$ torus, $(a_\tau,a_x,a_y)=(0,0,0)$, the partition function vanishes, 
\begin{align}
    Z^{\text{Majorana}}_{0,0,0}(g)=0,
    \label{eq:majorana_cone_ppp}
\end{align} due to the presence of a zero mode.  

In other cases, the Majorana cone partition function is the square root of the Dirac cone partition function, 
\begin{align}
    Z^{\text{Majorana}}_{a_\tau,a_x,a_y}(g)=\left|Z_{a_\tau,a_x,a_y}^{\text{Dirac}}(g)\right|^{\frac{1}{2}}.
    \label{eq:majorana_cone}
\end{align}
The Dirac cone partition functions have been shown to be modular covariant on $T^3$.\cite{Hsieh2016} In particular, under a modular transformation $U\in SL(3,\mathbb{Z})$, 
\begin{align}
    Z_{U(\mathbf{a})}(UgU^T)=Z_{\mathbf{a}}(g),
    \label{eq:fermion_modular_transformation}
\end{align}
where $\mathbf{a}\equiv (a_\tau,a_x,a_y)$. Under modular transformations, the $Z_{(0,0,0)}(g)$ sector is invariant, while other sectors rotated into each other. 

Due to the relation $(\ref{eq:majorana_cone})$, one can see that $Z^{\text{Majorana}}_{\mathbf{a}}(g)$ also follows the modular covariant condition (\ref{eq:fermion_modular_transformation}). And from (\ref{eq:majorana_cone_ppp}), one finds that $Z^{\text{Majorana}}_{0,0,0}(UgU^T)=Z^{\text{Majorana}}_{0,0,0}(g)$. 

Then it follows that the summation of all the components, 
\begin{align}
    Z(g)=\sum_{a_\tau,a_x,a_y=0,\frac{1}{2}} Z^{\text{Majorana}}_{a_\tau,a_x,a_y}(g),
\end{align}
is modular invariant $Z(UgU^T)=Z(g)$. The summation $Z(g)$ is nothing but $Z_m^{MC}$ in (\ref{eq:Z_f1form}), since $Z_{e,a_x,a_y}=\frac{1}{2}(Z_{0,a_x,a_y}+Z_{\frac{1}{2},a_x,a_y})$.

\section{Calculation of zero point energies}
\label{app:energies}

In this section, we give the details of our calculation of the zero-point energies of the $2+1$D Majorana system in different flux sectors.

First, let's do the calculation for the $1+1$D Majorana chain. Consider a chain of $2N$ real Majorana modes $\gamma_1$, ..., $\gamma_{2N}$, where $\{\gamma_i,\gamma_j\} = 2\delta_{ij}$. WLOG, we are going to assume that $N$ is even. The Hamiltonian is given by
\begin{equation}
H = \sum_{n=1}^N i (1+m) \gamma_{2n-1}\gamma_{2n} + i (1-m) \gamma_{2n}\gamma_{2n+1}
\end{equation}
where $\gamma_{2N+s} = \pm \gamma_s$ with $+$ sign in the $p$ sector and $-$ sign in the $ap$ sector. 

The Hamiltonian can be solved with Fourier transform
\begin{equation}
\tilde{\gamma}^e_k = \frac{1}{\sqrt{N}} \sum_{n=1}^N e^{ikn} \gamma_{2n}, \  \tilde{\gamma}^o_k = \frac{1}{\sqrt{N}} \sum_{n=1}^N e^{ikn} \gamma_{2n-1},
\end{equation}
where $k = 0, \pm \frac{2\pi}{N}, \pm 2\frac{2\pi}{N}$..., $\pm\left(\pi-\frac{2\pi}{N}\right),\pi$ to satisfy the boundary condition $\gamma_{2N+s} = \gamma_s$ and $k = \pm \frac{\pi}{N}, \pm 3\frac{\pi}{N}$..., $\pm\left(\pi-\frac{\pi}{N}\right)$ to satisfy the boundary condition $\gamma_{2N+s} = -\gamma_s$. $\tilde{\gamma}_{k=0}$ and $\tilde{\gamma}_{k=\pi}$ are real.
\begin{equation}
\left(\tilde{\gamma}_{k=0}\right)^2 = 1, \ \left(\tilde{\gamma}_{k=\pi}\right)^2 = 1
\end{equation}
When $k\neq 0$ or $\pi$, $\tilde{\gamma}_k$ is complex and $\tilde{\gamma}_{-k} = \tilde{\gamma}^{\dagger}_k$. 
\begin{equation}
\tilde{\gamma}_{k}\tilde{\gamma}_{-k} + \tilde{\gamma}_{-k}\tilde{\gamma}_{k} = 2
\end{equation}

The inverse Fourier transform is given by
\begin{equation}
\gamma_{2n} = \frac{1}{\sqrt{N}} \sum_{k} e^{-ikn} \tilde{\gamma}^e_{k}, \  \gamma_{2n-1} = \frac{1}{\sqrt{N}} \sum_{k} e^{-ikn} \tilde{\gamma}^o_{k}, 
\end{equation}

After the Fourier transform, the Hamiltonian becomes
\begin{equation}
\begin{array}{lll}
H & = &  2im \tilde{\gamma}^o_{k=0}\tilde{\gamma}^e_{k=0} + 2i\tilde{\gamma}^o_{k=\pi}\tilde{\gamma}^e_{k=\pi} \\
& & + \sum_{0<k<\pi} [-i(1+m)+i(1-m)e^{-ik}]\left(\tilde{\gamma}^e\right)^{\dagger}_{k}\tilde{\gamma}^o_k \\
& & + [i(1+m)-i(1-m)e^{ik}]\left(\tilde{\gamma}^o\right)^{\dagger}_k\tilde{\gamma}^e_{k} \\
&  = &  \epsilon_0 i\tilde{\gamma}^o_{k=0}\tilde{\gamma}^e_{k=0} + \epsilon_{\pi} i\tilde{\gamma}^o_{k=\pi}\tilde{\gamma}^e_{k=\pi} \\
&  & + \sum_{0<k<\pi} \epsilon_k \left(\tilde{\gamma}_{k,a}^{\dagger}\tilde{\gamma}_{k,a} - \tilde{\gamma}_{k,b}^{\dagger}\tilde{\gamma}_{k,b}\right)
\end{array}
\end{equation}
where
\begin{equation}
\epsilon_0 = 2m, \epsilon_{\pi} = 2, \epsilon_{0<k<\pi} = \sqrt{2(1+m^2)-2(1-m^2)\cos k}
\end{equation}
The $k=0$ and $k=\pi$ terms are only present in the $p$ sector. $\tilde{\gamma}_{k,a}$ and $\tilde{\gamma}_{k,b}$ are eigenmodes at momentum $k$. The zero-point energy in the two sectors are
\begin{equation}
\begin{array}{lll}
E^p & = & -\epsilon_0 -\epsilon_{\pi} - \sum_{0<k<\pi} 2 \epsilon_k \\
E^{ap} & = &  \sum_{0<k<\pi} 2\epsilon_k
\end{array}
\end{equation}
The factor of $2$ in front of $\epsilon_k$ accounts for the fact that $\{\tilde{\gamma}_k,\tilde{\gamma}_{-k}\}=2$. When $m=1$, we can check that the zero-point energy in both sectors are $-2N$, as it should be.

We can numerically evaluate $\Delta E = E^{p}-E^{ap}$ and see how it scales with $N$. Note that the sum over $0<k<\pi$ in the two sectors are over different $k$ points.  We see that when the chain is gapless, $\Delta E$ scales as $1/N$, when the chain is gapped, $\Delta E$ scales as $e^{-\alpha N}$. 

\begin{figure}[htbp]
\begin{center}
\includegraphics[width=3.3in]{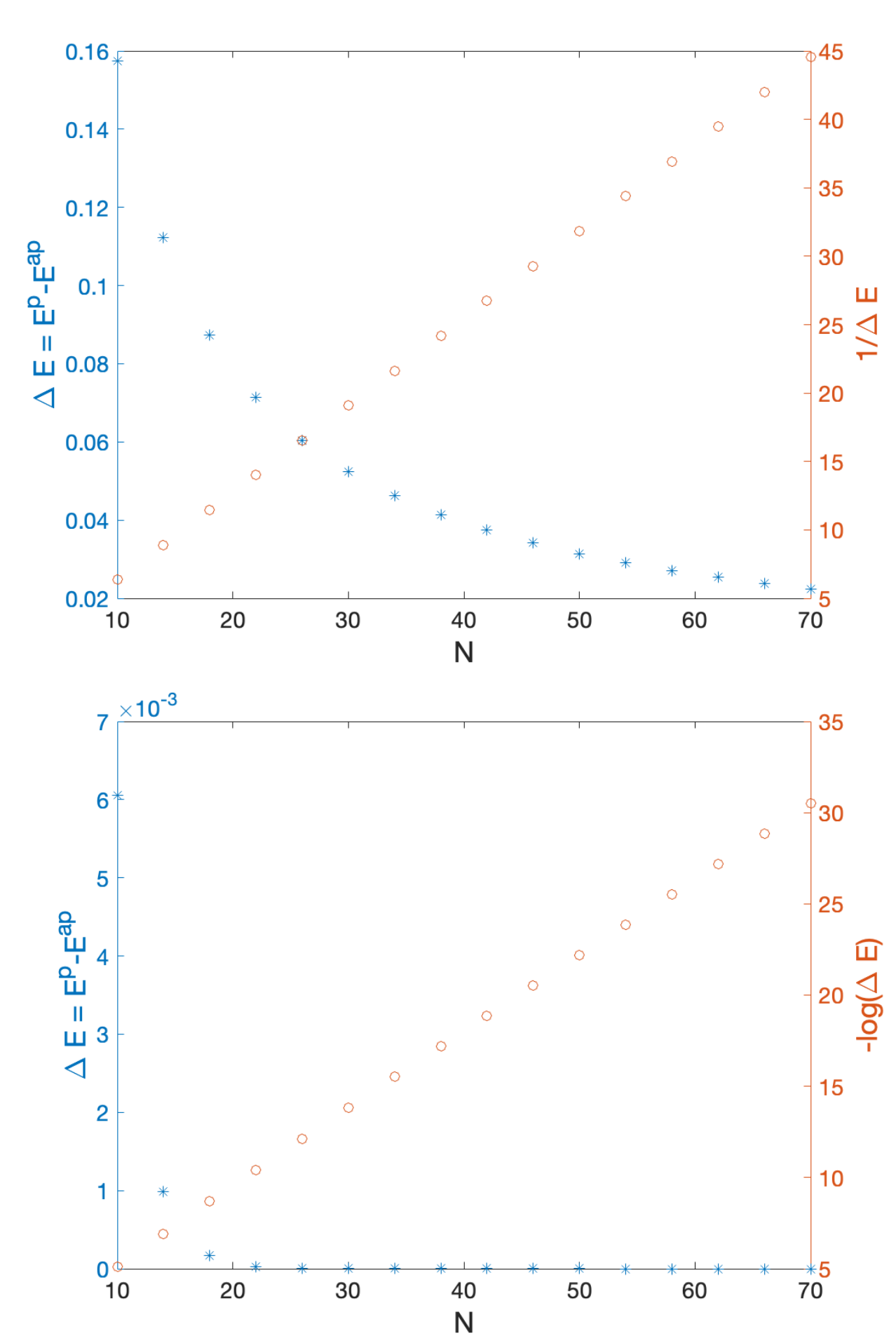}
\caption{Zero-point energy difference between the $p$ and $ap$ sectors in the $1+1$D Majorana chain. top: gapless, $\Delta E$ and $1/\Delta E$ vs $N$; bottom: gapped, $\Delta E$ and $-\log \Delta E$ vs $N$.} 
\label{fig:1DdE}
\end{center}
\end{figure}

If we normalize $\Delta E$ by the energy of the excited state in the $ap$ sector at momentum $\frac{4\pi}{N}$, we find that the ratio converges to $1/8$ which is exactly the total conformal dimension of the flux operator in the $1+1$D Ising CFT.
\begin{equation}
\Delta E \Big / \left(\frac{1}{2}\epsilon\left(\frac{4\pi}{N}\right)\right) \to 1/8
\end{equation}

Now let us do the calculation for the $2+1$D Majorana model. The Hamiltonian was given in Eq.~\ref{eq:2dH}
\begin{equation}
\begin{array}{ll}
H =  \sum_{m,n} & -t \left(c^{\dagger}_{m+1,n}c_{m,n} + c^{\dagger}_{m,n+1}c_{m,n} +h.c. \right)\\
& - (\mu-4t) c^{\dagger}_{m,n}c_{m,n} \\  & + \left(\Delta c^{\dagger}_{m+1,n}c^{\dagger}_{m,n} + i\Delta c^{\dagger}_{m,n+1}c^{\dagger}_{m,n}+h.c. \right)
\end{array}
\end{equation}

Take Fourier transform
\begin{equation}
\begin{array}{lll}
c_{k_x,k_y} & = & \frac{1}{\sqrt{N_x}} \frac{1}{\sqrt{N_y}} \sum_{m,n} e^{ik_xm}e^{ik_yn} c_{m,n}, \\ c^{\dagger}_{k_x,k_y} & = & \frac{1}{\sqrt{N_x}} \frac{1}{\sqrt{N_y}} \sum_{m,n} e^{-ik_xm}e^{-ik_yn} c^{\dagger}_{m,n}
\end{array}
\end{equation}
such that $\{c^{\dagger}_{k_x,k_y},c_{k_x,k_y}\} = 1$.

The inverse transform is given by
\begin{equation}
\begin{array}{lll}
c_{m,n} & = & \frac{1}{\sqrt{N_x}} \frac{1}{\sqrt{N_y}} \sum_{k_x,k_y} e^{-ik_xm}e^{-ik_yn} c_{k_x,k_y}, \\ c^{\dagger}_{m,n} & = & \frac{1}{\sqrt{N_x}} \frac{1}{\sqrt{N_y}} \sum_{k_x,k_y} e^{ik_xm}e^{ik_yn} c^{\dagger}_{k_x,k_y}
\end{array}
\end{equation}
The Hamiltonian after the transformation is given by
\begin{equation}
\begin{array}{lll}
H & = & \sum_{k_x,k_y} (4t-\mu-2t\cos k_x -2t\cos k_y) c^{\dagger}_{k_x,k_y} c_{k_x,k_y} \\
 & & + \Delta \left((e^{ik_x}+ie^{ik_y})c^{\dagger}_{k_x,k_y}c^{\dagger}_{-k_x,-k_y} + h.c. \right)
\end{array}
\end{equation}

In the $(p,p)$ sector, at $k_x = 0$ or $\pi$ and $k_y = 0$ or $\pi$, there is one complex mode at each point with energy  $2t(2-\cos k_x - \cos k_y)$. As  $2t(2-\cos k_x - \cos k_y)\geq 0$, these modes do not contribute to zero point energy.

At $k_x,k_y \neq 0, \pi$, $(k_x,k_y)$ and $(-k_x,-k_y)$ label different modes. Set $k_x>0$ or $k_x=0$, $k_y>0$. The two modes mix together as
\begin{equation}
\begin{array}{l}
a(k_x,k_y) + \\
\begin{pmatrix} c^{\dagger}_{k_x,k_y} & c_{-k_x,-k_y} \end{pmatrix}.  \begin{pmatrix} a(k_x,k_y) & b(k_x,k_y)\\ b^*(k_x,k_y) &  -a(k_x,k_y) \end{pmatrix}   \begin{pmatrix} c_{k_x,k_y} \\ c^{\dagger}_{-k_x,-k_y} \end{pmatrix}  
\end{array}
\end{equation} 
where $a(k_x,k_y) = 4t-\mu-2t\cos k_x -2t\cos k_y$, $b(k_x,k_y) = \Delta (2i\sin k_x-2\sin k_y)$. Therefore, the two modes at $(k_x,k_y)$ and $(-k_x,-k_y)$ mix into two complex modes with energy
\begin{equation}
\epsilon(k_x,k_y) = \pm \sqrt{a^2(k_x,k_y)+4\Delta^2 \sin^2 k_y + 4\Delta^2 \sin^2 k_x}
\end{equation}
The total contribution to zero point energy is hence
\begin{equation}
E(k_x,k_y) = -\sqrt{a^2(k_x,k_y)+4\Delta^2 \sin^2 k_y + 4\Delta^2 \sin^2 k_x} + a(k_x,k_y)
\end{equation}
This is consistent with the fact that the zero point energy at $k_x,k_y = 0, \pi$ is zero. Therefore, if we take a sum of $E(k_x,k_y)$ over all momentum points, we get twice the zero point energy. 

\begin{figure}[H]
\begin{center}
\includegraphics[width=3.3in]{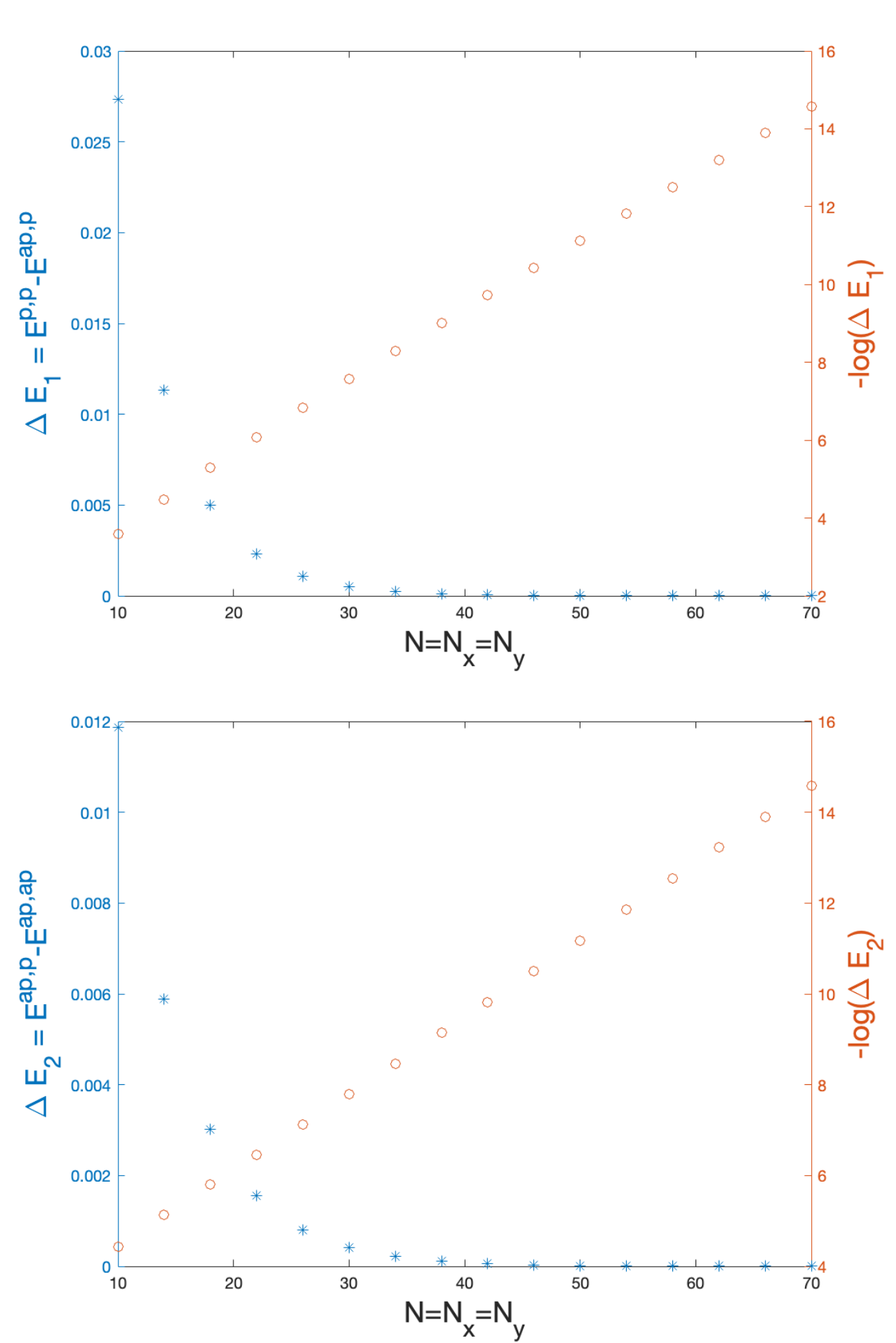}
\caption{Zero-point energy difference between the different flux sectors in the $2+1$D gapped Majorana model. top: the energy difference between the $(p,p)$ and $(ap,p)$ sectors, $\Delta E$ and $-\log(\Delta) E$ vs $N$; bottom: the energy difference between the $(ap,p)$ and $(ap,ap)$ sectors, $\Delta E$ and $-\log(\Delta) E$ vs $N$.} 
\label{fig:2DdEgap}
\end{center}
\end{figure}

\begin{figure}[H]
\begin{center}
\includegraphics[width=3.3in]{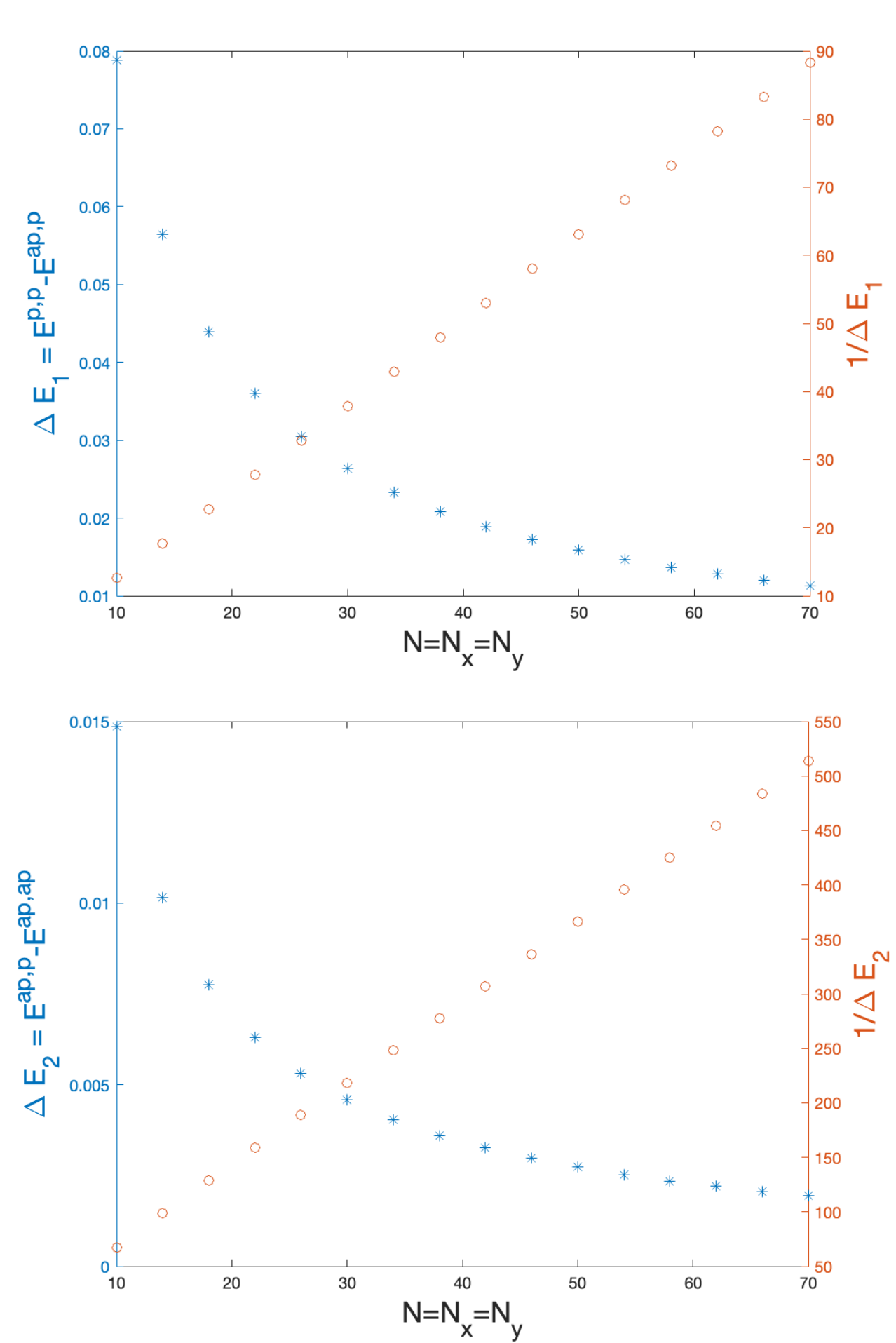}
\caption{Zero-point energy difference between the different flux sectors in the $2+1$D gapless Majorana model. top: the energy difference between the $(p,p)$ and $(ap,p)$ sectors, $\Delta E$ and $1/\Delta E$ vs $N$; bottom: the energy difference between the $(ap,p)$ and $(ap,ap)$ sectors, $\Delta E$ and $1/\Delta E$ vs $N$.} 
\label{fig:2DdEgapless}
\end{center}
\end{figure}

Now we can calculate the energy difference in different sectors by summing $E(k_x,k_y)$ over the corresponding momentum points and taking the difference. We see that, similar to $1+1$D, when the system is gapless, the energy difference scales as $1/N$, when the system is gapped, the energy difference scales as $e^{-\alpha N}$. The plots correspond to parameter choice $t=1$, $\Delta=0.5$, $\mu=0$ (gapless) or $\mu=-0.2$ (gapped). But we have verified that the scaling behavior does not change when the parameters are varied slightly. 

In the gapless case, we can normalize the energy difference and find the coefficient of the $1/N$ scaling. These numbers are listed in Table~\ref{table:scaling}.

\section{Lattice derivation of symmetry action induced by Cheshire defect}
\label{sec:VC}

In this section, we are going to derive the quantum channel operation in Eq.~\ref{eq:VC} induced by sweeping a Cheshire string defect using explicit lattice calculation. We are going to do the derivation using $2+1$D Toric Code. The result naturally generalizes to $3+1$D Toric Code.

As discussed in Ref.~\cite{Tantivasadakarn2024}, Cheshire strings can be generated with a 1D sequential circuit and then moved around and fused using 1D finite-depth circuits. We are going to demonstrate the whole process of creating two Cheshire strings, moving them apart, sweeping through the whole system, and finally bringing them back together and annihilating them using quantum circuits. In particular, we are going to see that the process involves two single-qubit measurements and therefore corresponds to a quantum channel operation on the logical space.

\begin{figure}[ht]
    \centering
    \includegraphics[scale=0.45]{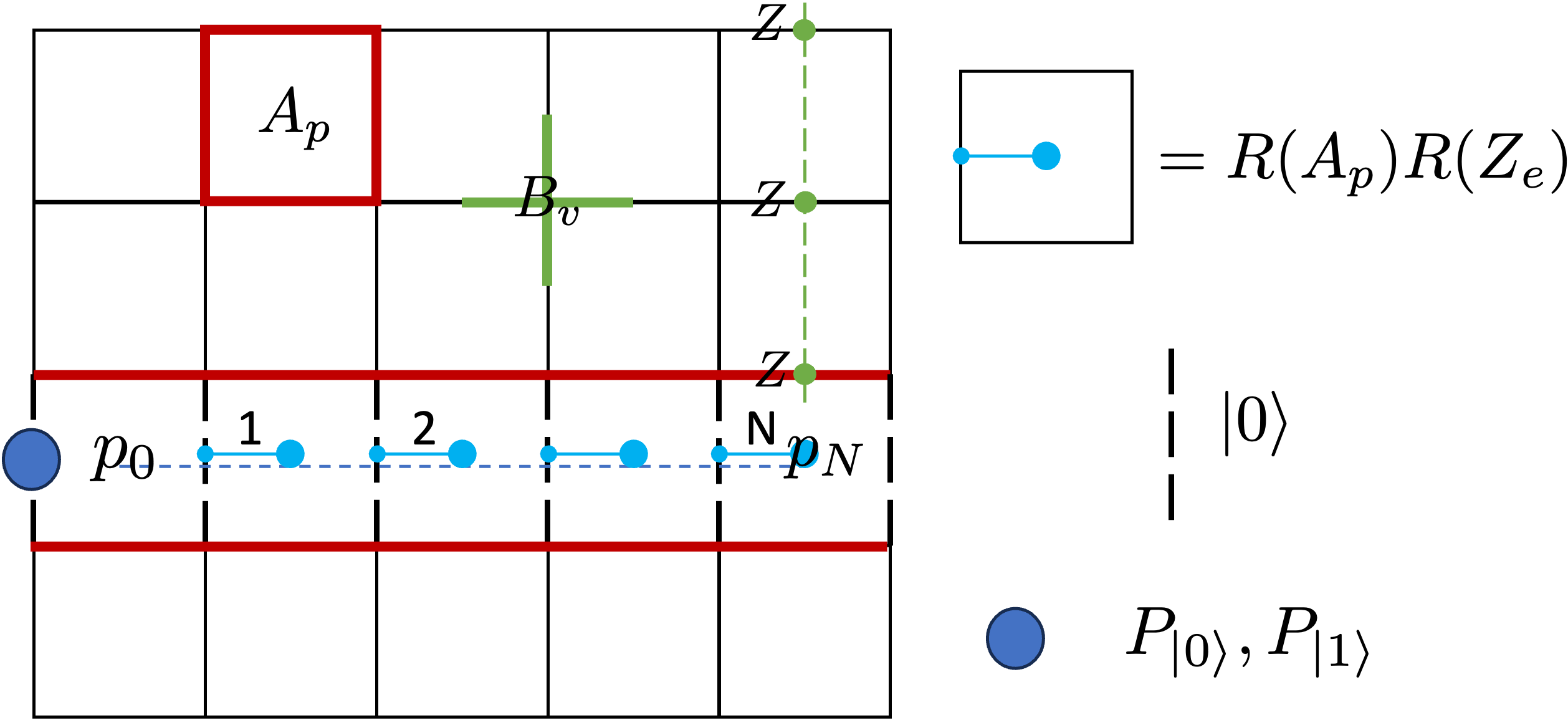}
    \caption{Generation of an $e$-Cheshire string in $2+1$D Toric Code with a sequential circuit. }
    \label{fig:2DSe}
\end{figure}

Fig.~\ref{fig:2DSe} shows the generation of an $e$-Cheshire string using a sequential quantum circuit in the $2+1$D Toric Code. $A_p = \prod_{e\in p} X_e$ and $B_v = \prod_{v \in e} Z_e$ are Hamiltonian terms of the $2+1$D Toric Code. A Cheshire string (on the dual lattice) from $p_0$ to $p_N$ is generated by applying a sequence of gate sets represented by the blue dot pairs. Here
$R(\mathcal O) \equiv  e^{-\frac{i\pi}{4} \mathcal O}$, which has the property that for Pauli operators $P$ and $Q$,
\begin{align}
    R(Q) P R(Q)^\dagger = \begin{cases}
    P ;& [P,Q]=0\\
    iPQ; & \{P,Q \}=0
    \end{cases}
\end{align}
The dashed black edges are mapped to the product state $|0\rangle$ forming the condensate. To complete the condensation across the nontrivial $x$-direction cycle (whose total charge is measured by $\prod X_e$ along the two red lines), we need to map the last edge on the boundary also to a product $|0\rangle$ or $|1\rangle$ state. This is achieved with a measurement in the $Z$ basis on the last edge indicated by the big dark blue dot. The measurement projects the last edge into either the $|0\rangle$ or the $|1\rangle$ state which actually directy corresponds to the eigenvalue of $W^x_e$, the $e$ string operator in the $x$-direction. To see the correspondence, notice that $W^x_e$, which is the product of $Z_e$'s along a dual string in the $x$-direction, commutes with all the unitary gates in the sequential circuit. Therefore its eigenvalue is preserved. Along the particular dual string along the Cheshire defect, after the sequential circuit all but one edge are mapped to the $|0\rangle$ state. Therefore, the last edge is in the $|0\rangle$ state if $W^x_e = 1$ and in the $|1\rangle$ state if $W^x_e = -1$. That is, to generate a Cheshire string along the non-trivial cycle in the $x$-direction, we need to measure the $W^x_e$ operator and project to its eigensectors. This is the first part of the quantum channel. 


\begin{figure}[ht]
    \centering
    \includegraphics[scale=0.45]{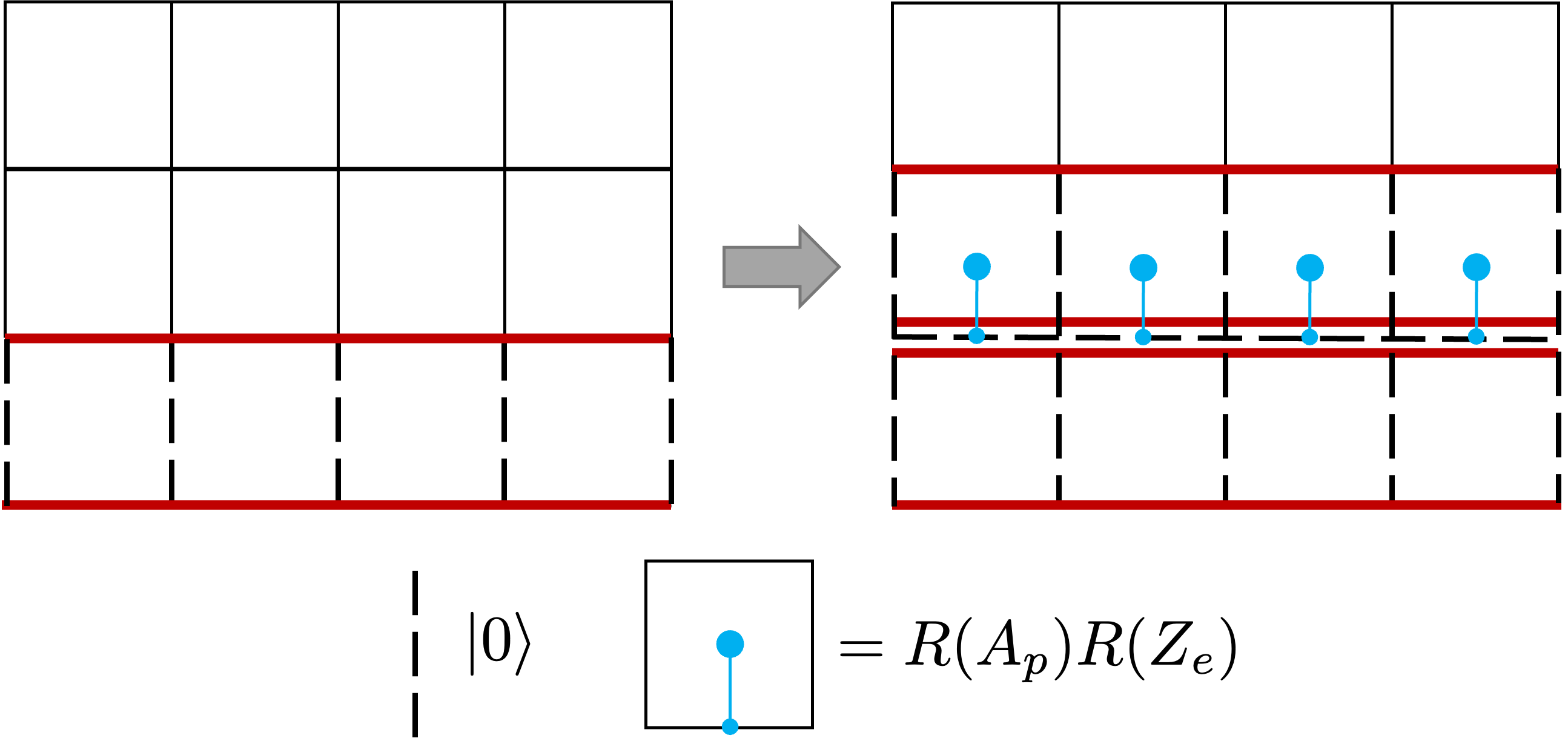}
    \caption{Splitting one Cheshire string into two strings with coherent tunneling in between.}
    \label{fig:2DSe1-2}
\end{figure}

In the second step, we split the single defect into two. This can be achieved using the circuit shown in Fig.~\ref{fig:2DSe1-2} where gate sets in each plaquette are applied in parallel (because they commute with each other). The ``fattened'' Cheshire string can be thought of as two strings, each with width one. The edges in between the two strings are in the $|0\rangle$ state -- the ground state of $Z_e$ on these edges, indicating the coherent tunneling of charges between the two strings. The total charge of the two strings (measured by $\prod X_e$ along the top and bottom red lines) is conserved, while the total charge of each string (measured by $\prod X_e$ along the top two red lines and the bottom two lines) are not individually conserved.

\begin{figure}[ht]
    \centering
    \includegraphics[scale=0.45]{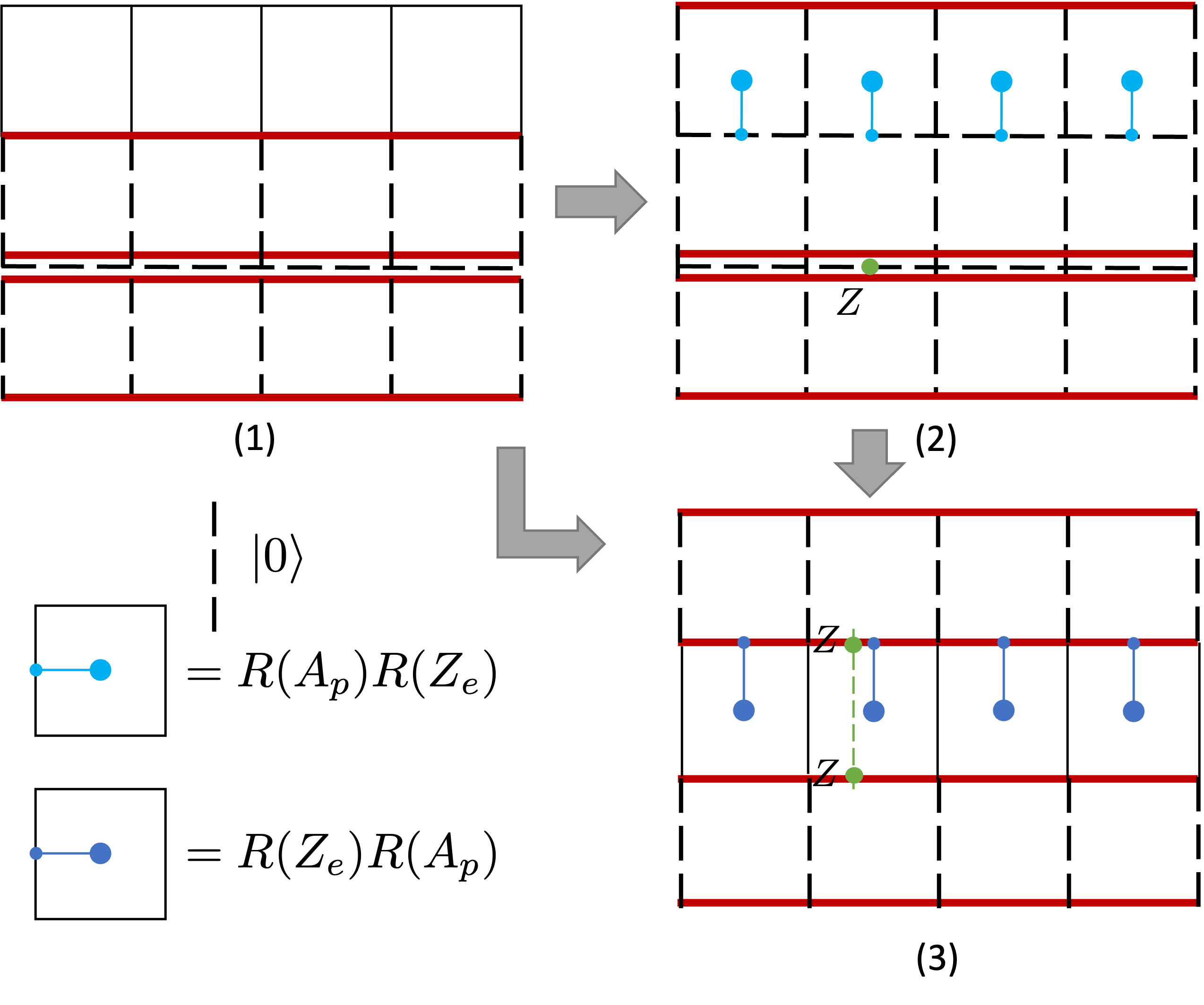}
    \caption{Moving one Cheshire string away from the other while maintaining the coherent tunneling in between.}
    \label{fig:2DSe_move}
\end{figure}

\begin{figure}[ht]
    \centering
    \includegraphics[scale=0.45]{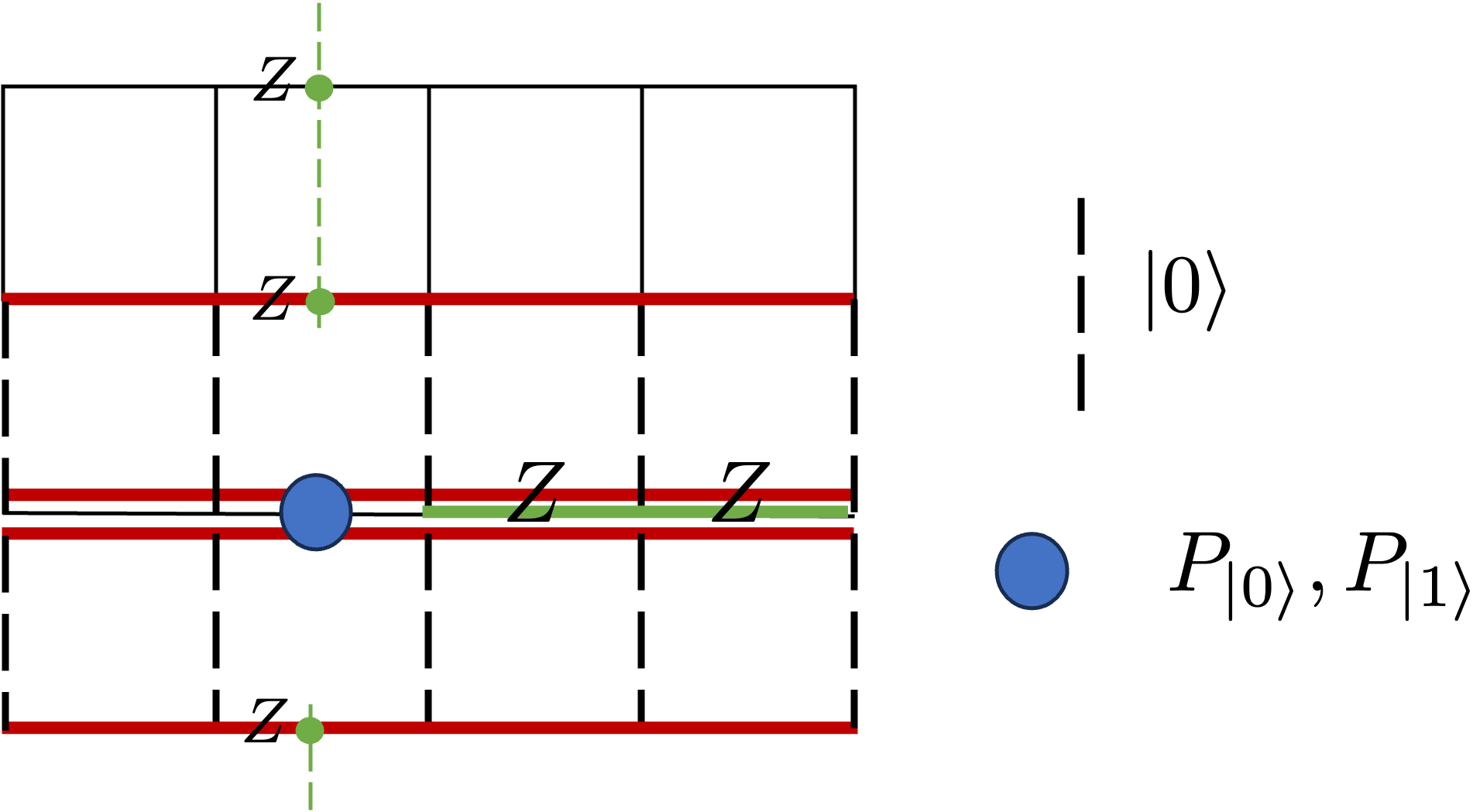}
    \caption{Fusing two Cheshire strings back together into one.}
    \label{fig:2DSe_fuse}
\end{figure}

Then we can move the two strings apart using the circuit shown in Fig.~\ref{fig:2DSe_move}. After the two steps shown in the figure, the top string is moved up by one lattice spacing. The coherent tunneling between the two strings, which was originally enforced by a single $Z_e$ (the green dot in (2)), is now enforced by the $Z_eZ_e$ term (the pair of green dots in (3)). We can repeat this step $L_y-2$ number of times so that the two strings come back together as shown in Fig.~\ref{fig:2DSe_fuse} and the coherent tunneling is enforced by a long string of $\prod Z_e$ between them. 

When the two Cheshire strings are brought back together as shown in Fig.~\ref{fig:2DSe_fuse}, the edges on the line between them are in the ground state of pairwise $Z_eZ_e$ interaction (the solid green line). This line is the $\mathcal{Z}_2$ coefficient in fusing two Cheshire strings and carries a two fold degeneracy. To fuse the two strings together, we can remove the degeneracy by projecting one of the edges in the $Z_e$ basis (big blue dot). Once we project one edge in either $|0\rangle$ or $|1\rangle$ state, all the other edges follow and the two Cheshire strings are fused together. The projection of a single edge into the $|0\rangle$ or $|1\rangle$ state is directly related to the eigenvalue of $W^y_e$. To see this, notice that $W^y_e$ commutes with all previous steps and hence remains invariant. Because the coherent tunneling between the two Cheshire strings is enforced, the product of all but one edges in $W^y_e$ is equal to $1$. Therefore, the last edge will project to $|0\rangle$ if $W^y_e = 1$ and will project to $|1\rangle$ if $W^y_e = -1$. Therefore, to fuse two Cheshire strings back together, we need to measure $W^y_e$ and project to its eigensectors. This is the second part of the quantum channel. 

Finally, we can annihilate the remaining Cheshire string by reserving the procedure shown in Fig.~\ref{fig:2DSe}. Note that we can skip the measurement step because we are already in an eigensector of $W^x_e$. Now putting the whole process together, we see that the symmetry operation corresponding to sweeping a Cheshire string defect is a projection onto the common eigenstates of $W^x_e$ and $W^y_e$. If all four measurement results are allowed, we get the quantum channel described in Eq.~\ref{eq:VC}. 

Let us comment on the relation of our result to the fusion table of Toric Code defects given in Ref.~\cite{Roumpedakis2023}. For the four-dimensional ground space of the Toric Code on the 2D torus, we will choose a basis such that string operators act as logical operator on the two logical qubits as
\begin{equation}
W^x_e \sim Z_1, \ W^y_e \sim Z_2, \ W^x_m \sim X_2, \ W^y_m \sim X_1.
\end{equation}
In this basis, the logical operation corresponding to the $e$-Cheshire string $S_e$ is a projection in computation basis $|00\rangle$, $|01\rangle$, $|10\rangle$, $|11\rangle$. If we post-select on the projection result, we get four different operators
\begin{equation}
\begin{array}{ll}
V^{00}_e \sim |00\rangle\langle 00|, & V^{01}_e \sim |01\rangle\langle 01|, \\
V^{10}_e \sim |10\rangle\langle 10|, & V^{11}_e \sim |11\rangle\langle 11|.
\end{array}
\end{equation}
Similarly, we can conclude that the logical operation corresponding to the $m$-Cheshire string $S_m$ is a projection into basis $|++\rangle$, $|+-\rangle$, $|-+\rangle$, $|--\rangle$ and if we post-select, we get
\begin{equation}
\begin{array}{ll}
V^{++}_m \sim |++\rangle\langle ++|, & V^{+-}_e \sim |+-\rangle\langle +-|, \\
V^{-+}_e \sim |-+\rangle\langle -+|, & V^{--}_e \sim |--\rangle\langle --|.
\end{array}
\end{equation}
The defect $S_{\psi}$ exchanges $e$ and $m$ and hence maps between $W^x_e$ and $W^x_m$ as well as $W^y_e$ and $W^y_m$. It's action on the logical space is hence
\begin{equation}
V_{\psi} \sim \text{SWAP} \left(H \otimes H\right),
\end{equation}
where $\text{SWAP}$ exchanges the two logical qubits and $H$ is the Hadamard gate. Note that this is consistent with our interpretation in section~\ref{sec:fZ2} that the Majorana chain defect corresponds to a controlled-$Z$ logical operation in the common eigen-basis of $W^x_{f}$ and $W^y_{f}$.

Now we can see how the fusion rule of the logical operations can be consistent with the fusion rule of the defects. To see the correspondence, we need to pick one element out of four for $V_e$ and one out of four for $V_m$. For example, we can pick $V^{00}_e$ and $V^{++}_m$ and observe that their fusion rule together with $V_{\psi}$ match exactly with that of $S_e$, $S_m$ and $S_{\psi}$ in Ref.~\cite{Roumpedakis2023}. We list a few of the fusion results here (fusion coefficients are ignored)
\begin{align}
V^{00}_eV^{00}_e  & = & |00\rangle\langle 00| & = & V^{00}, \\
V_{\psi}V^{00}_eV_{\psi} & = & |++\rangle \langle ++| & = & V^{++}_m, \\
V^{00}_eV^{++}_m & = & |00\rangle \langle ++| & = & V^{00}_eV_{\psi}, \\
V^{\psi}V^{\psi} & = & I. & & 
\end{align}
Similar fusions are satisfied if we choose pairs $V^{01}_e$ and $V^{-+}_m$, $V^{10}_e$ and $V^{+-}_m$, $V^{11}_e$ and $V^{--}_m$.

\end{document}